\documentclass[sigconf,balance=false]{acmart}

\AtBeginDocument{%
  }

\usepackage{algorithm}
\usepackage{algorithmic}
\usepackage{amsmath}

\usepackage{amssymb}
\usepackage{booktabs}
\usepackage{multirow}
\usepackage{graphicx}
\usepackage{xcolor}
\usepackage{subcaption}
\usepackage{enumitem}
\usepackage{pifont}
\makeatletter
\let\@fnsymbol@acmorig\@fnsymbol
\def\@fnsymbol#1{\ifcase#1\or\raisebox{-0.15ex}{\ding{41}}\else\@fnsymbol@acmorig{#1}\fi}
\makeatother

\newcommand{\beatedit}{\textsc{BeatEdit}}
\newcommand{\beat}{\textsc{Beat}}
\newcommand{\tagkeep}{\texttt{KEEP}}
\newcommand{\tagdelete}{\texttt{DELETE}}
\newcommand{\tagreplace}{\texttt{REPLACE}}
\newcommand{\tagappend}{\texttt{APPEND}}
\newcommand{\tagshift}{\texttt{SHIFT}}

\copyrightyear{2026}
\acmYear{2026}
\setcopyright{none}
\acmConference[MM '26]{Proceedings of the 34th ACM International Conference on Multimedia}{November 10--14, 2026}{Rio de Janeiro, Brazil}
\acmBooktitle{Proceedings of the 34th ACM International Conference on Multimedia (MM '26), November 10--14, 2026, Rio de Janeiro, Brazil}
\acmDOI{10.1145/3767308.3835589}
\let\BeatEditCopyrightFootnote\footnotetextcopyrightpermission
\renewcommand\footnotetextcopyrightpermission[1]{%
  \BeatEditCopyrightFootnote{%
    \href{https://creativecommons.org/licenses/by/4.0/}{%
      \includegraphics[height=5ex]{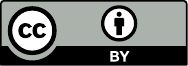}}\par
    ACM publication license: Creative Commons Attribution 4.0 International.\par
    \textcopyright{} 2026 Copyright held by the owner/author(s).\par
    This is an extended author's version, including Appendices A--P.
    Accepted for publication in \emph{Proceedings of the 34th ACM International
    Conference on Multimedia (MM '26)}, November 10--14, 2026,
    Rio de Janeiro, Brazil.\par
    The definitive Version of Record will be published by ACM at
    \url{https://doi.org/10.1145/3767308.3835589}.%
  }%
}

\begin{document}

%% ============================================================
%% TITLE
%% ============================================================
\title{BeatEdit: Symbolic Music Generation as Explicit Editing}

\author{Haoyu Gu}
\orcid{0009-0005-2597-0366}
\affiliation{%
  \department{School of Future Technology}
  \institution{South China University of Technology}
  \city{Guangzhou}
  \country{China}
}
\email{ghy20050104@gmail.com}

\author{Lekai Qian}
\orcid{0009-0003-1307-7939}
\affiliation{%
  \department{School of Future Technology}
  \institution{South China University of Technology}
  \city{Guangzhou}
  \country{China}
}
\email{ftqlk@mail.scut.edu.cn}

\author{Haowu Zhou}
\orcid{0009-0005-6195-0688}
\affiliation{%
  \department{School of Future Technology}
  \institution{South China University of Technology}
  \city{Guangzhou}
  \country{China}
}
\email{202364870491@mail.scut.edu.cn}

\author{Qi Liu}
\orcid{0000-0001-5378-6404}
\authornote{Corresponding authors.}
\affiliation{%
  \department{School of Future Technology}
  \institution{South China University of Technology}
  \city{Guangzhou}
  \country{China}
}
\email{drliuqi@scut.edu.cn}

\author{Shuai Wang}
\orcid{0000-0003-1523-9631}
\authornotemark[1]
\affiliation{%
  \department{School of Intelligence Science and Technology}
  \institution{Nanjing University}
  \city{Suzhou}
  \country{China}
}
\email{shuaiwang@nju.edu.cn}

\renewcommand{\shortauthors}{Haoyu Gu, Lekai Qian, Haowu Zhou, Qi Liu, and Shuai Wang}

%% ============================================================
%% ABSTRACT
%% ============================================================
\begin{abstract}
Music creation is fundamentally a process of revision. Yet symbolic music generation remains dominated by paradigms that produce complete sequences from scratch, with limited support for selective modification. Edit-based methods have proven effective for text transformation tasks, but remain largely unexplored for symbolic music. We trace this absence to the representational level: conventional event-based music encodings lack the structural properties required by explicit music editing. In contrast, the \beat{} encoding, a beat-grid-anchored representation originally designed for autoregressive generation, possesses structural properties amenable to editing. We propose \beatedit{}, the first framework for symbolic music generation based on explicit edit operations, recasting generation as producing new content by editing a draft rather than synthesizing from scratch. \beatedit{} comprises three complementary mechanisms along an axis of increasing edit density: per-token sequence tagging for error correction, iterative refinement for accompaniment editing, and tag-then-fill for segment completion. All these mechanisms share a single encoding and pre-trained backbone, achieving higher precision and perceptual quality than autoregressive and diffusion methods across all three tasks, while remaining efficient, with single-pass inference completing in under 100\,ms. Cross-encoding evaluation further reveals that encoding design substantially influences editing effectiveness, with notable encoding–method interaction effects. Code is available at \url{https://github.com/Haoyu-Gu/BeatEdit-code}.
\end{abstract}



\maketitle

\section{Introduction}\label{sec:intro}

%% ¶1: Music creation as revision
In music production, much of the creative effort lies in refining what already exists rather than writing from scratch. A performer corrects wrong notes, an arranger reshapes accompaniment without altering the melody, and a composer fills in missing bars while preserving surrounding material. These tasks share a common nature: input and output overlap substantially, and the work is to locate what needs change and modify it, not to regenerate the whole.

%% ¶2: Current paradigms contradict revision
Symbolic music generation, however, is dominated by autoregressive~\cite{huang2018music,payne2019musenet,lu2023musecoco,vonrutte2023figaro} and diffusion~\cite{min2023diffusion,huang2024ruleguided} models, neither of which natively supports selective modification: autoregressive generation must regenerate from the edit point onward even for a single-note change, and diffusion does not explicitly model which positions require change. Works that add local modification atop these paradigms---music inpainting~\cite{pati2019inpainting}, SDEdit-style noise-then-denoise~\cite{meng2022sdedit}, discrete diffusion conditioning~\cite{lv2023getmusic}---still establish no explicit operational semantics at the symbolic sequence level, and remain variants of implicit generation.

%% ¶3: NLP editing success raises the question
In natural language processing, edit-based methods have proven effective for grammatical error correction~\cite{omelianchuk2020gector,awasthi2019parallel}, sentence rewriting~\cite{malmi2019lasertagger,gu2019levenshtein}, and controlled text generation~\cite{mallinson2020felix,raheja2023coedit}: an explicit edit operation (keep, delete, replace, or insert) is predicted for each input position, structurally ensuring that unedited content is preserved. This differs from masked infilling and autoregressive completion in mechanism: the editor itself localizes the positions to modify and applies typed operations there, whereas conditional generation requires the target region to be specified externally. Whether the three music tasks above---which span a wide range of edit density---admit the same formulation is the question we take up, and answering it calls for a unified framework covering all three regimes.

%% ¶4: Representational barrier
Yet these methods remain largely unexplored for symbolic music. The obstacle lies at the
representational level. In text, words serve as natural atomic units,
source and target sequences align position-by-position before and
after editing, and a single-word modification does not cascade to
neighboring positions. Existing symbolic music encodings lack these
properties: in REMI~\cite{huang2020remi}, for instance, a single
note is scattered across four interleaved tokens, any insertion or
deletion shifts all subsequent positions, and the flat event stream
provides no structural boundary to contain edit propagation. We formalize this gap as three structural requirements that edit-based
methods impose on the underlying encoding
(Section~\ref{sec:encoding}), and show that the \beat{}
encoding~\cite{beat2025}, a beat-grid-anchored representation,
naturally satisfies them. Its only residual dependency---cascading
relative pitch offsets within a beat---is addressed by mechanism-level
adaptations (Section~\ref{sec:method}).

%% ¶5: BeatEdit framework
Building on these structural properties, we propose \beatedit{}, a framework for symbolic music generation based on explicit edit operations; throughout, ``generation'' means producing new content by editing a draft, not synthesizing from scratch. \beatedit{} implements three complementary mechanisms along an axis of increasing edit density: per-token sequence tagging for sparse point edits, iterative deletion-insertion-prediction for progressive voice-level refinement, and a two-stage tag-then-fill pipeline for segment completion. All three share the same beat-level encoding and pre-trained backbone, with dedicated adaptations for music-specific properties absent in text (\S\ref{sec:related}).

%% ¶6: Contributions
\begin{enumerate}[nosep,leftmargin=*]
\item \textbf{Paradigm.} We propose \beatedit{}, the first framework that formulates symbolic music generation as explicit editing, with three complementary mechanisms spanning the edit-density axis under a single pre-trained backbone.
\item \textbf{Representational analysis.} We formalize three structural requirements (R1--R3) that edit-based methods impose on the underlying encoding. A $2\times2$ encoding--method factorial reveals that encoding choice has a statistically significant impact on performance, with strong encoding--method interaction effects, establishing it as a previously overlooked design lever.
\item \textbf{Empirical results.} On 192K piano pieces, edit-based
    methods nearly double the best generative baseline in exact-match
    correction with single-pass inference under 100\,ms, two orders of magnitude faster than
    autoregressive generation. Ablations further show that
    domain-specific BERT pre-training is the critical performance
    driver.
\end{enumerate}

%% ============================================================
%% 2. EDIT-COMPATIBLE MUSIC ENCODING
%% ============================================================
\section{Edit-Compatible Music Encoding}\label{sec:encoding}

%% ---- Figure 1: Encoding Pipeline ----
\begin{figure*}[t]
  \centering
  \includegraphics[width=0.99\textwidth]{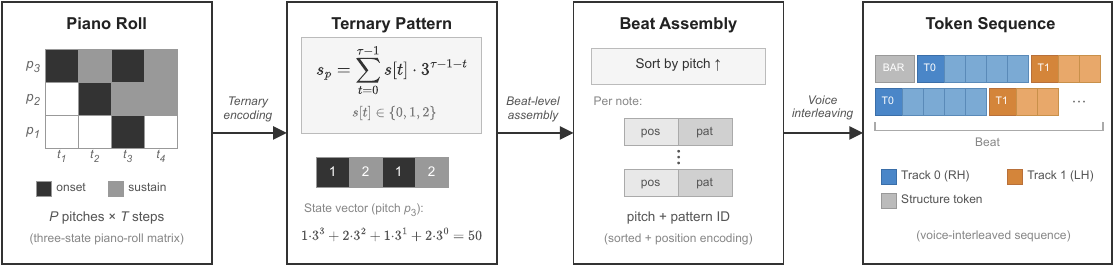}
  \caption{Encoding pipeline. Each pitch becomes a pattern token $s_p$ (rhythm) and a position token $d$ (pitch). Beats from both voices are interleaved with voice markers.}
  \label{fig:encoding}
\end{figure*}

\subsection{Design Requirements}\label{sec:requirements}

Edit-based methods operate by assigning exactly one edit label to each input position. In text, this mechanism works seamlessly because words naturally serve as atomic units with stable positional correspondence. Transferring this paradigm to symbolic music requires the underlying encoding to satisfy three analogous properties:

\begin{itemize}[nosep,leftmargin=*]
  \item[\textbf{R1}] \emph{Atomic edit units.} Each note should map to a minimal, self-contained token group, so that one edit label corresponds to one complete editing target.
  \item[\textbf{R2}] \emph{Trivial source--target alignment.} Source and target sequences must align position-by-position, so that edit labels can be extracted by simple comparison.
  \item[\textbf{R3}] \emph{Edit locality.} A single-note musical edit should affect only a bounded number of tokens in the sequence; the result must remain a valid encoding without requiring global re-computation.
\end{itemize}

\begin{table}[t]
\caption{Edit-compatibility of these existing encodings (\ding{51}\,=\,satisfied, \ding{55}\,=\,violated, $\sim$\,=\,partial).}
\label{tab:encoding_req}
\centering
\small
\begin{tabular}{@{}lcccl@{}}
\toprule
& \textbf{R1} & \textbf{R2} & \textbf{R3} & \textbf{Bottleneck} \\
\midrule
MIDI-Like~\cite{oore2020feeling}
  & \ding{55} & \ding{55} & \ding{55}
  & 3--4 tok/note; onset/offset decoupled \\
REMI~\cite{huang2020remi}
  & \ding{55} & \ding{55} & \ding{55}
  & 4 tok/note; insert/delete shifts all \\
Structured~\cite{hadjeres2021piano}
  & $\sim$ & \ding{55} & $\sim$
  & 4 tok/note; relative time propagates \\
CPWord~\cite{hsiao2021compound}
  & $\sim$ & \ding{55} & $\sim$
  & Factored heads; no structural grid \\
\midrule
\beat{}~\cite{beat2025} (Rel.)
  & \ding{51} & \ding{51} & $\sim$
  & Rel.\ pitch cascades within beat \\
\beat{}~\cite{beat2025} (Abs.)
  & \ding{51} & \ding{51} & \ding{51}
  & --- \\
\bottomrule
\end{tabular}
\end{table}

As summarized in Table~\ref{tab:encoding_req}, existing symbolic music encodings have difficulty satisfying R1--R3 simultaneously. MIDI-Like~\cite{oore2020feeling} and REMI~\cite{huang2020remi} both represent each note with 3--4 interleaved tokens, scattering onset and duration across non-adjacent positions and preventing self-contained edit units (R1); without a fixed temporal grid, any insertion or deletion shifts all subsequent positions, breaking positional alignment (R2) and allowing edits to propagate unboundedly (R3). Structured~\cite{hadjeres2021piano} and CPWord~\cite{hsiao2021compound} achieve partial atomicity (R1) but lack a structural grid that confines edits locally: the former relies on relative time-shifts, causing duration modifications to propagate to subsequent positions (R2, R3); the latter decomposes attributes into factored prediction heads, whose cross-attribute coupling hinders independent per-position label assignment (R2). Octuple~\cite{zeng2021musicbert} and MMT~\cite{dong2023mmt} couple multiple dimensions into composite representations or interleaved tracks, so that editing a single note involves coordinated changes across multi-dimensional tokens (R1, R3). These failure modes point to a common structural need: a fixed temporal grid that partitions the sequence into content-invariant alignment units, maps each note to a compact token group, and bounds edit propagation locally.

\subsection{\beat{} Encoding}\label{sec:ternary}
\beat{}~\cite{beat2025} introduced ternary pattern encoding and relative pitch compression for autoregressive music generation. Its structural properties naturally align with the editing requirements above, and we therefore adopt it as the underlying representation for \beatedit{}. Below we describe the encoding mechanism and show how it satisfies R1 and R2; its satisfaction of R3 and the residual cascading dependency are discussed in Section~\ref{sec:variants}. We adopt a dual-voice piano setting (right hand and left hand) with ascending pitch order, explicit boundary markers, and no velocity channel. These choices, and the fixed $\tau{=}4$ resolution below, are constraints of our controlled evaluation rather than of the editing paradigm: boundary markers can be supplied by standard beat-tracking tools at deployment, and velocity is an orthogonal pattern channel edited by the same operations. The encoding converts a piano roll into a compact token sequence through three steps, illustrated in Figure~\ref{fig:encoding}: encoding each pitch's within-beat rhythm as a pattern token, assembling active pitches into a beat-level sequence, and interleaving beats across voices. Intuitively, the piano roll is cut into beat-sized columns; within each column, every active pitch's rhythm is compressed into a single base-3 digit string, and the resulting per-beat token groups are laid out in temporal order. Token types and ID ranges are detailed in Appendix~\ref{app:encoding_spec}.

\noindent\textbf{Step 1: Pattern encoding.} We represent each piano piece as a three-state piano roll $X \in \{0,1,2\}^{P \times T}$, where $P$ is the number of pitches and $T$ the number of time steps, quantized at sixteenth-note resolution ($\tau = 4$ steps per beat). Entry $X[p,t]$ is 1 if a note at pitch $p$ has its onset at step $t$, 2 if that note is held through step $t$, and 0 if pitch $p$ is silent at step $t$. Within each beat starting at step $t_0$, the state vector $\mathbf{s}_p = \left(X[p,t_0], \ldots, X[p,t_0{+}\tau{-}1]\right) \in \{0,1,2\}^\tau$ of pitch $p$ is compressed into a single \emph{pattern token} via base-3 conversion, $s_p = \sum_{t=0}^{\tau-1} \mathbf{s}_p[t]\,3^{\tau-1-t}$, yielding $3^\tau = 81$ possible values. Each pattern token is an atomic unit encoding the complete within-beat rhythm of one pitch. For example, a quarter note occupies the whole beat as $\mathbf{s}_p = (1,2,2,2)$ and maps to $s_p = 1{\cdot}3^3 + 2{\cdot}3^2 + 2{\cdot}3^1 + 2{\cdot}3^0 = 53$; two eighth notes $(1,2,1,2)$ map to $s_p = 50$.

\noindent\textbf{Step 2: Beat-level assembly.} To eliminate sparsity, only active pitches are encoded. For each beat, active pitches are sorted in ascending order $p^{(1)} < \cdots < p^{(M)}$ and encoded using relative offsets\label{eq:relpos}: $d^{(1)} = p^{(1)}$, and $d^{(j)} = p^{(j)} - p^{(j-1)}$ for $j > 1$. Each note becomes exactly two tokens, a position token $d^{(j)}$ and a pattern token $s^{(j)} := s_{p^{(j)}}$, i.e., the pattern token of the $j$-th active pitch as defined in Step~1, forming a self-contained edit unit that satisfies R1. Empty beats are represented by a single marker token.

\noindent\textbf{Step 3: Voice interleaving.} Within each bar, beats from both voices are interleaved in a fixed order: melody and accompaniment alternate at every beat position. Because the temporal axis is partitioned into fixed-length beats, every beat occupies a dedicated token span regardless of its content. This ensures that the $n$-th beat of the source maps to the $n$-th beat of the target regardless of local edits, providing trivial alignment (R2). Furthermore, each beat is a self-contained subsequence delimited by voice markers, so edits within one beat do not affect tokens outside that beat, satisfying edit locality (R3). Although we describe dual-voice interleaving for piano, the same mechanism generalizes to multi-track settings (Section~\ref{sec:multi_instrument}). The resulting vocabulary contains 185 tokens; a typical piano piece encodes to 1{,}200--2{,}500 tokens under separated encoding.

\subsection{Encoding Variants}\label{sec:variants}
The encoding described above follows the original \beat{} design, which uses relative pitch offsets and separate position and pattern tokens. While effective for autoregressive generation, editing introduces \emph{cascading dependencies}: under relative encoding, modifying one note's pitch forces the next note's position token to change as well, even though that note is itself correct. This cascade is bounded within a single beat (preserving R3) but motivates exploring alternative encoding choices. We organize the design space along two orthogonal axes, yielding a $2\times2$ family of schemes in which the bundled variants (C, D) are introduced in this work (Table~\ref{tab:scheme_comparison}):

\begin{table}[t]
\caption{Encoding scheme design space ($2 \times 2$).}
\label{tab:scheme_comparison}
\centering
\begin{tabular}{@{}llcccc@{}}
\toprule
& & \textbf{Tok/note} & \textbf{Vocab} & \textbf{Labels} & \textbf{Avg.\ len} \\
\midrule
A & Abs + Sep & 2 & 186 & 350 & 1{,}907 \\
B & Rel + Sep & 2 & 185 & 350 & 1{,}891 \\
C & Rel + Bun & 1 & 7{,}145 & 14{,}258 & 1{,}163 \\
D & Abs + Bun & 1 & 7{,}145 & 14{,}258 & 1{,}163 \\
\bottomrule
\end{tabular}
\end{table}

The first axis is \emph{position encoding}. The relative scheme (\S\ref{sec:ternary}, Step~2) exploits music's translational invariance but introduces cascading dependencies, which are handled by the \tagshift{} mechanism (Section~\ref{sec:edit_ops}). The absolute scheme encodes each note's pitch as a direct index, eliminating cascading at the cost of sacrificing translational invariance. As we show in Section~\ref{sec:experiments}, cascading errors compound across refinement rounds, making relative encoding particularly problematic for multi-pass editing methods.

The second axis is \emph{token organization}. The separated scheme represents each note as two tokens (position + pattern), keeping the label space compact. The bundled scheme packs both into a single token, reducing sequence length (Table~\ref{tab:scheme_comparison}) but expanding the vocabulary and label space substantially.

All four schemes share the same beat-level alignment and can be combined with any editing method. The optimal scheme depends on the task--method interaction, as we show in Section~\ref{sec:experiments}.

%% ============================================================
%% 3. THE \beat{}EDIT FRAMEWORK
%% ============================================================
\section{The BeatEdit Framework}\label{sec:method}

%% ---- Figure 2: BeatEdit Framework Overview ----
\begin{figure*}[t]
  \centering
  \includegraphics[width=.99\textwidth]{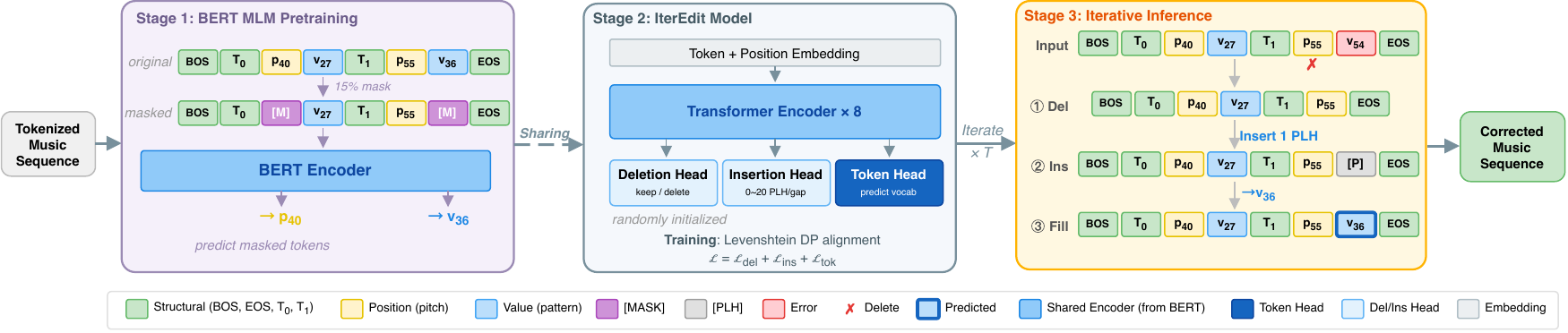}
  \caption{IterEdit pipeline. Stage~1: BERT MLM pre-training. Stage~2: three heads trained jointly with the encoder. Stage~3: iterative delete--insert--fill until convergence.}
  \label{fig:framework}
\end{figure*}

With an encoding that satisfies the structural requirements for editing in place, we now turn to the mechanisms that operate on it.
\beatedit{} implements three complementary editing mechanisms, sequence tagging (\S\ref{sec:error_correction}), iterative refinement (\S\ref{sec:task3}), and tag-then-fill (\S\ref{sec:task2}), all sharing a pre-trained backbone (\S\ref{sec:bert}). Because all methods operate on the same representation, each can be applied across tasks beyond its primary assignment, a property we exploit in Section~\ref{sec:experiments}.

\subsection{Shared Pre-trained Backbone}\label{sec:bert}

The \beat{} encoding assigns every beat a self-contained token span, allowing a single encoder to serve all three tasks within each encoding scheme. We use a BERT~\cite{devlin2019bert} encoder (8~layers, 512~hidden, ${\sim}$26.6M parameters), pre-trained via masked language modeling on 192K piano pieces. All three editing methods share this encoder as their backbone, differing only in the task-specific heads attached on top. During pre-training, only music tokens are masked; structural tokens (beat boundaries, bar markers) remain visible so that the beat grid required by R2 is preserved throughout. Training is augmented with random transposition ($\pm 5$ semitones). Ablation experiments (\S\ref{sec:discussion}) confirm that this pre-training step is the critical performance driver: removing it leads to substantial degradation across all tasks (Appendix~\ref{app:ablation}).

\subsection{Sequence Tagging}\label{sec:error_correction}

For sparse point edits, a per-token label can preserve or transform each position without length-changing generation. We extend standard text edit operations (keep, delete, replace, append) with a music-specific $\tagshift{}_{\pm n}$\label{sec:edit_ops} operation that handles the cascading pitch dependencies unique to relative encoding (Section~\ref{sec:variants}). Given a position token $d_i$ representing a relative pitch offset, \tagshift{} applies an integer correction $\tagshift{}_{\pm n}(d_i) = d_i \pm n$ with $n \in \{1,\dots,5\}$. This is necessary because modifying a note's pitch under relative encoding shifts all subsequent offsets within the beat: correcting note $j$ from $p_j$ to $p_j'$ requires simultaneously applying $\tagshift{}_{p_j - p_j'}$ to $d_{j+1}$ to maintain the correct absolute pitch of the unchanged note $j{+}1$. Under bundled encoding, each note is a single token and pitch corrections use $\tagreplace{}$ over the full vocabulary instead. The complete label space comprises 350 labels for separated encoding and 14{,}258 for bundled encoding (Appendix~\ref{app:labels}). Edit labels are extracted via beat-level alignment, since perturbations preserve beat count.

Prediction is decomposed into two linear heads over the shared encoder's per-position representation: a binary \emph{error detection head} $\hat{e}_i$ predicting whether each token requires editing, and a \emph{tag prediction head} $\hat{t}_i$ predicting the specific edit operation (formulations in Appendix~\ref{app:implementation}). At inference, the detection head gates tag predictions: $\hat{t}_i$ is overridden to \tagkeep{} wherever $\hat{e}_i < \theta$, providing a conservative bias against unnecessary edits.

Training pairs are constructed synthetically with fresh random perturbations each epoch (Appendix~\ref{app:perturbation}). The combined loss is $\mathcal{L} = \mathcal{L}_{\text{tag}} + \lambda \cdot \mathcal{L}_{\text{detect}}$, where $\mathcal{L}_{\text{tag}}$ is weighted cross-entropy with down-weighted \tagkeep{} to counteract the $>$85\% keep-dominance, and $\mathcal{L}_{\text{detect}}$ is binary cross-entropy. Training proceeds with a frozen encoder followed by full fine-tuning with differential learning rates. At inference, the model iterates up to three rounds with a \emph{decode--filter--reencode}\label{sec:inference} post-processing cycle to ensure output validity (Algorithm~3 in Appendix~\ref{app:seqtag_alg}; walkthrough in Appendix~\ref{app:walkthrough}).

\subsection{Iterative Refinement}\label{sec:task3}

When edits become denser and span many beats---as in accompaniment editing, where tokens must be inserted, deleted, and replaced throughout the accompaniment while preserving the melody---single-pass tagging cannot adjust sequence length. We address this regime with an iterative deletion-insertion-prediction approach. The encoding's beat-level voice interleaving makes accompaniment separation trivial via positional indexing.

Three task-specific heads are attached to the shared pre-trained encoder (\S\ref{sec:bert}). We adopt two key adaptations for the music domain. First, an \emph{encoder-only} architecture provides bidirectional context over both melody and partially-refined accompaniment, which we find essential for editing tasks (Section~\ref{sec:discussion}). Second, \emph{per-token editable flags} $E_i \in \{0, 1\}$ enforce hard melody preservation: all melody tokens and structural markers are set to $E_i = 0$, and the model is constrained to never modify these positions at both training and inference time.

Three linear heads share the encoder representation (Appendix~\ref{app:implementation}): a \emph{deletion head} predicting per-token deletion probability, an \emph{insertion head} predicting from each adjacent token pair how many placeholders to insert in the gap between them (up to 20), and a \emph{token head} that fills all placeholders. The deletion head only operates where $E_i{=}1$; melody tokens are never deleted. The joint training loss is $\mathcal{L}_{\text{iter}} = \mathcal{L}_{\text{del}} + \mathcal{L}_{\text{ins}} + \mathcal{L}_{\text{tok}}$, all three terms being cross-entropy losses. Token prediction uses label smoothing ($\epsilon{=}0.1$) to encourage diversity.

Training uses a four-level perturbation strategy ranging from light single-note edits (L1) to near-complete accompaniment rewriting (L4), with \emph{intermediate-state sampling}: rather than always training from the fully perturbed input, we sample intermediate states by partially applying ground-truth edits, reducing exposure bias between training and multi-round inference. At inference, each iteration executes deletion, insertion, and prediction in sequence; convergence is reached when a round produces zero changes (Algorithm~4 in Appendix~\ref{app:iteredit_alg}; hyperparameters in Appendix~\ref{app:implementation}).

\subsection{Tag-then-Fill}\label{sec:task2}

At the far end of the edit-density axis, segment completion requires reconstructing a contiguous span of missing beats from context alone. Because no source tokens exist within the gap, neither tagging nor iterative refinement can operate; we therefore adopt a two-stage decomposition into structural allocation (tagging) and content generation (filling). Both stages use the shared pre-trained encoder (\S\ref{sec:bert}).

\noindent\textbf{Stage 1: Tagging.} A tagger assigns each token $x_i$ one of 11~labels from the set $\{\tagkeep{},\; \tagdelete{},\; \tagreplace{}_{\texttt{M}},\; \tagappend{}_{\texttt{M} \times k}\}$ where $k \in \{1,\dots,8\}$, allocating space for missing content. $\tagreplace{}_{\texttt{M}}$ substitutes the token with a single \texttt{[MASK]}, and $\tagappend{}_{\texttt{M} \times k}$ inserts $k$ masks after the token. The tagger is trained with Focal Loss~\cite{lin2017focal} ($\gamma{=}2.0$) to handle the strong class imbalance toward \tagkeep{} (Appendix~\ref{app:implementation}). The tagger's output is deterministically converted into a \emph{skeleton sequence} $\tilde{X}$ with \texttt{[MASK]} placeholders at positions requiring new content.

\noindent\textbf{Stage 2: Filling.} An inserter predicts tokens at all mask positions. Rather than filling all masks simultaneously, we adopt an iterative confidence-ranked schedule over $T{=}2$ passes. At each pass $t$, the inserter predicts distributions over the vocabulary for all remaining masks; the top $\lfloor|\mathcal{M}| / (T - t + 1) \rfloor$ positions by confidence are accepted and committed as $\hat{x}_i = \arg\max_v P_{\text{ins}}(v \mid \tilde{X})$ for $i \in \mathcal{M}_t$, where $\mathcal{M}_t$ is the set of unfilled mask positions at pass $t$. Only mask positions are gathered for prediction, keeping computation proportional to the gap size.

The tagger's output serves as a hard structural constraint: the inserter can only modify positions explicitly marked for filling, so any errors are confined to the gap region by construction. Training details are provided in Appendix~\ref{app:implementation}.

%% ============================================================
%% 4. EXPERIMENTS
%% ============================================================
\section{Experiments}\label{sec:experiments}

We evaluate \beatedit{} across all three tasks using a unified evaluation framework. We first validate \beat{}'s structural advantage over alternative encodings (\S\ref{sec:encoding_comparison}), then present unified main results (\S\ref{sec:main_results}), analyze encoding--method interactions (\S\ref{sec:encoding_effect}), compare against an external diffusion baseline (\S\ref{sec:boundary}), report multi-instrument generalization (\S\ref{sec:multi_instrument}), and discuss implications (\S\ref{sec:discussion}).

\subsection{Experimental Setup}\label{sec:setup}

\textbf{Dataset.} 192,788 piano pieces from the MuseScore Collection, split 80\%/10\%/10\% train/validation/test at the song level. Sequences exceeding 2,048 tokens are cropped at bar boundaries. Since paired draft-to-final corpora do not yet exist for music---a field-wide gap---we synthesize edit pairs by corrupting clean targets, as is standard in edit-based NLP~\cite{omelianchuk2020gector,awasthi2019parallel}; this also lets us control task difficulty precisely. Code, preprocessing scripts, and configurations are released at \url{https://github.com/Haoyu-Gu/BeatEdit-code}.

\textbf{Encoding schemes.} A $2 \times 2$ factorial design over \emph{position encoding}---Absolute vs.\ Relative---and \emph{token organization}---Separated vs.\ Bundled---yields four schemes: A~(Abs+Sep), B~(Rel+Sep), C~(Rel+Bun), D~(Abs+Bun); see Table~\ref{tab:scheme_comparison}. For external comparison, we also evaluate MIDI-Like~\cite{oore2020feeling}, Structured~\cite{hadjeres2021piano}, REMI~\cite{huang2020remi}, and CPWord~\cite{hsiao2021compound} with the same SeqTag architecture.

\textbf{Methods.} Three editing methods: \textbf{SeqTag}, a sequence tagger for error correction~(\S\ref{sec:error_correction}); \textbf{IterEdit}, an iterative refinement model for accompaniment editing~(\S\ref{sec:task3}); and \textbf{TagFill}, a tag-then-fill pipeline for segment completion~(\S\ref{sec:task2}). SeqTag and TagFill share the same pre-trained backbone. SeqTag is evaluated on correction and editing. IterEdit is trained for editing and correction using the multi-level perturbation strategy of \S\ref{sec:task3} and evaluated across all four schemes; a separate contiguous-masking configuration is used for completion. TagFill is evaluated on all three tasks.

\textbf{Baselines.} No-Edit returns the input unchanged; Copy-Ctx copies a random context beat to the corrupted region; BERT-CMLM re-predicts all positions via masked language modeling. We include two families of autoregressive (AR) baselines. \emph{AR}: for correction and editing, a ${\sim}$162M-parameter causal LLM (LLaMA) trained from scratch on the same data, evaluated under four strategies with increasing positional guidance (Prompt, Teacher, Detect, Selective; Appendix~\ref{app:llama}); the main table reports AR-Detect, which uses SeqTag's error detector for beat-level localization. For completion, we use Anticipatory~\cite{thickstun2024anticipatory}, a ${\sim}$128M-parameter AR model with bidirectional infilling, evaluated with its publicly available pre-trained weights. Diffusion uses D3PM~\cite{austin2021d3pm} discrete diffusion with ${\sim}$34M parameters and SDEdit-style noise-then-denoise at start ratio $r{=}0.3$; Appendix~\ref{app:diffusion} reports $r{=}0.5, 0.7$.

\textbf{Evaluation.} 3~tasks $\times$ 4~encodings $\times$ 200~samples = 2{,}400 test instances, with 95\% bootstrap CIs for key comparisons. Difficulty levels are distributed as L1\,:\,L2\,:\,L3\,:\,L4 = 30\,:\,30\,:\,25\,:\,15. Four metrics are reported: beat\_exact\_match~($\uparrow$), the fraction of perturbed beats with exactly matching tokens; note\_f1~($\uparrow$), note-level F1; MPE~($\downarrow$), mean pitch error in semitones; and FMD~($\downarrow$), Fr\'{e}chet Music Distance~\cite{retkowski2024fmd} on full sequences. All metrics except FMD are computed on perturbed beats only; full per-scheme breakdowns are reported in Appendix~\ref{app:full_results}. Note that beat\_exact\_match and note\_f1 reward only recovery of the specific ground truth---appropriate for correction, but blind to plausible alternatives; Section~\ref{sec:boundary} therefore adds distribution-overlap metrics.

\subsection{Cross-Encoding Comparison}\label{sec:encoding_comparison}

\begin{table}[t]
\caption{Cross-encoding comparison.}
\label{tab:beat_vs_others}
\centering
\small
\begin{tabular}{@{}l@{\hspace{8pt}}c@{\hspace{8pt}}c@{\hspace{8pt}}c@{\hspace{8pt}}c@{}}
\toprule
& \textbf{tok/note} & \textbf{nf1} & \textbf{recovery} & \textbf{MPE}$\downarrow$ \\
\midrule
MIDI-Like$^\dagger$          & 3.75 & 0.860 & $-$13.1\% & ---   \\
Structured                    & 4.08 & 0.882 & 1.3\%     & 0.205 \\
CPWord                        & 1.56 & 0.891 & 8.4\%     & 0.192 \\
REMI                          & 3.59 & 0.893 & 10.0\%    & 0.183 \\
\midrule
\beat{} (Rel, Sep)           & 2.61 & 0.923 & 34.1\%    & 0.141 \\
\beat{} (Rel, Bun)           & 1.59 & 0.927 & 37.2\%    & 0.135 \\
\beat{} (Abs, Bun)           & 1.59 & 0.938 & 46.6\%    & 0.102 \\
\beat{} (Abs, Sep)           & 2.66 & \textbf{0.940} & \textbf{48.1\%} & \textbf{0.087} \\
\bottomrule
\multicolumn{5}{@{}l@{}}{\footnotesize $^\dagger$BERT MLM converged to higher perplexity (PPL\,=\,8.18 vs.\ ${\sim}$1.5 for others).}
\end{tabular}
\end{table}

All eight encodings share the same SeqTag architecture, training data, and perturbation protocol, with No-Edit baselines at note\_f1 $\approx 0.881$--$0.884$ confirming comparable corruption difficulty. The results separate into three tiers. At the bottom, MIDI-Like~\cite{oore2020feeling} falls below the No-Edit baseline, as its flat event stream prevents effective MLM pre-training (PPL\,=\,8.18 vs.\ ${\sim}$1.5 for others). Structured~\cite{hadjeres2021piano}, CPWord~\cite{hsiao2021compound}, and REMI~\cite{huang2020remi} form a middle tier with marginal recovery. All four \beat{} variants constitute the top tier, with the best scheme (Abs\,Sep) reaching $4.8\times$ higher recovery than the strongest middle-tier encoding and absolute encoding consistently outperforming relative.

The comparison with CPWord is particularly instructive: both are equally compact (${\sim}$1.6 tok/note), yet \beat{} achieves substantially higher recovery. The key difference is structural---\beat{}'s beat-level grid confines edits locally, whereas CPWord's factored heads introduce cross-attribute dependencies. Since all configurations share the identical architecture, the performance gap is entirely attributable to encoding design.

%% ---- Main Results Table ----
\begin{table*}[t]
  \caption{Main results (best scheme per method, perturbed beats only).}
  \label{tab:main_beat}
  \centering
  \begin{tabular}{@{}l ccc ccc ccc@{}}
  \toprule
  & \multicolumn{3}{c}{\textbf{Error Correction}} & \multicolumn{3}{c}{\textbf{Accomp.\ Editing}} & \multicolumn{3}{c}{\textbf{Segment Completion}} \\
  \cmidrule(lr){2-4} \cmidrule(lr){5-7} \cmidrule(lr){8-10}
  \textbf{Method} & beat$\uparrow$ & nf1$\uparrow$ & FMD$\downarrow$ & beat$\uparrow$ & nf1$\uparrow$ & FMD$\downarrow$ & beat$\uparrow$ & nf1$\uparrow$ & FMD$\downarrow$ \\
  \midrule
  SeqTag & \textbf{.726}\textsuperscript{B} & \textbf{.855} & \textbf{0.15} & .408\textsuperscript{C} & .574 & 2.56 & \multicolumn{3}{c}{N/A \,\,} \\
  IterEdit & .719\textsuperscript{A} & .849 & 1.03 & \textbf{.480}\textsuperscript{D} & \textbf{.607} & \textbf{1.19} & .120\textsuperscript{A} & .212 & 0.34 \\
  TagFill & .368\textsuperscript{A} & .667 & 2.03 & .350\textsuperscript{D} & .519 & 1.24 & \textbf{.436}\textsuperscript{A} & \textbf{.557} & \textbf{0.27} \\
  \midrule
  No-Edit & .030 & .644 & 2.69 & .037 & .473 & 2.48 & .000 & .000 & 1.76 \\
  Copy-Ctx & .129 & .429 & 7.74 & .127 & .368 & 3.80 & .116 & .032 & 2.62 \\
  BERT-CMLM & .388\textsuperscript{D} & .659 & 1.01 & .188\textsuperscript{D} & .456 & 0.97 & .111\textsuperscript{A} & .042 & 2.22 \\
  Diffusion & .390\textsuperscript{D} & .602 & 0.87 & .200\textsuperscript{C} & .372 & 2.75 & .080\textsuperscript{D} & .228 & ${>}$10 \\
  AR$^\dagger$ & .394\textsuperscript{A} & .650 & 1.24 & .278\textsuperscript{A} & .478 & 1.73 & .069 & .027 & 5.69 \\
  \bottomrule
  \end{tabular}
  \\[2pt]
  {\footnotesize $^\dagger$Correction \& editing: AR-Detect (LLaMA 162M + SeqTag error detector; Appendix~\ref{app:llama}). Completion: Anticipatory~\cite{thickstun2024anticipatory} (128M, pre-trained, bidirectional infilling).}
\end{table*}

\subsection{Main Results}\label{sec:main_results}

Table~\ref{tab:main_beat} summarizes the best configuration per method--task combination across all encoding schemes.

For \textbf{error correction}, SeqTag and IterEdit achieve near-identical beat exact match (0.726 vs.\ 0.719), both nearly doubling the best generative baselines ($\leq$0.394). The advantage is consistent across all metrics (Table~\ref{tab:main_beat}). TagFill trails at 0.368, indicating that its two-stage decomposition introduces cascade errors on a task where most edits are single-token replacements.

For \textbf{accompaniment editing}, IterEdit leads at 0.480, surpassing SeqTag (0.408) and TagFill (0.350), with the lowest FMD among editing methods. The per-difficulty breakdown (Table~\ref{tab:encoding_effect}b) reveals a crossover: SeqTag dominates at L1 where most edits are single-token replacements amenable to one-pass tagging, while IterEdit's iterative loop overtakes at L2--L4 as multi-token insertions and deletions become frequent. At L4 all methods drop sharply to below 0.10; we locate this boundary precisely in \S\ref{sec:discussion}.

\textbf{Segment completion} pushes edit density to its extreme: the entire gap contains no source tokens. SeqTag is inapplicable here as it has no generation head. TagFill dominates (beat: 0.436), reflecting the natural alignment between its masked-token inserter and the contiguous-gap structure. IterEdit achieves only 0.120, confirming that iterative refinement is less suited to generation from empty spans. Among baselines, Anticipatory~\cite{thickstun2024anticipatory}---an AR model designed for music infilling---achieves low token-level precision (beat: 0.069), illustrating the gap between generating plausible music and recovering specific target content. All other baselines also fall well below TagFill.

Across all three tasks, editing methods outperform generative baselines. The AR-Detect variant---which pairs SeqTag's error detector with autoregressive regeneration---trails editing methods by large margins despite using identical beat-level localization, confirming that the advantage lies in the edit mechanism itself rather than in error localization alone.

We further conducted a subjective study with 33 evaluators rating outputs on a 1--5 MOS scale over music quality, naturalness, and edit accuracy (Appendix~\ref{app:subjective}). The results track the objective metrics: \beat{} leads the other encodings by 1.4--1.8 MOS under the same SeqTag architecture, edit-based methods outscore generative baselines on correction (SeqTag 4.21, IterEdit 4.13 vs.\ Diffusion/CMLM 3.25, AR-Detect 3.05), and each mechanism leads its own task (TagFill 4.15 on completion, IterEdit 4.07 on editing).

\subsection{Encoding $\times$ Method Analysis}\label{sec:encoding_effect}

\begin{table*}[t]
  \caption{Beat exact match: \textbf{(a)}~by encoding scheme, \textbf{(b)}~by difficulty level.}
  \label{tab:encoding_effect}
  \centering
  \begin{minipage}[t]{0.48\textwidth}
  \centering
  \textbf{(a) Encoding $\times$ Method}\\[4pt]
  \begin{tabular}{@{}llcccc@{}}
  \toprule
  \textbf{Task} & \textbf{Method} & \textbf{A} & \textbf{B} & \textbf{C} & \textbf{D} \\
  \midrule
  \multirow{3}{*}{Corr.} & SeqTag & .700 & \textbf{.726} & .702 & .700 \\
                          & IterEdit   & \textbf{.719} & .070 & .412 & .569 \\
                          & TagFill  & \textbf{.368} & .061 & .335 & .329 \\
  \midrule
  \multirow{3}{*}{Edit.} & SeqTag & .382 & .394 & \textbf{.408} & .399 \\
                          & IterEdit   & .458 & .069 & .352 & \textbf{.480} \\
                          & TagFill  & .321 & .089 & .301 & \textbf{.350} \\
  \midrule
  \multirow{2}{*}{Compl.} & IterEdit   & \textbf{.120} & .060 & .011 & .005 \\
                          & TagFill  & \textbf{.436} & .264 & .298 & .326 \\
  \bottomrule
  \end{tabular}
  \end{minipage}
  \hfill
  \begin{minipage}[t]{0.48\textwidth}
  \centering
  \textbf{(b) Difficulty Level}\\[4pt]
  \begin{tabular}{@{}llcccc@{}}
  \toprule
  \textbf{Task} & \textbf{Method} & \textbf{L1} & \textbf{L2} & \textbf{L3} & \textbf{L4} \\
  \midrule
  \multirow{3}{*}{Corr.} & SeqTag\textsuperscript{B} & .757 & .737 & \textbf{.700} & \textbf{.690} \\
                          & IterEdit\textsuperscript{A}   & \textbf{.790} & \textbf{.743} & .647 & .664 \\
                          & TagFill\textsuperscript{A}  & .407 & .421 & .256 & .379 \\
  \midrule
  \multirow{3}{*}{Edit.} & SeqTag\textsuperscript{C} & \textbf{.773} & .415 & .208 & .045 \\
                          & IterEdit\textsuperscript{D}   & .763 & \textbf{.583} & \textbf{.298} & \textbf{.090} \\
                          & TagFill\textsuperscript{D}  & .477 & .442 & .286 & .076 \\
  \midrule
  \multirow{2}{*}{Compl.} & IterEdit\textsuperscript{A}   & .168 & .112 & .108 & .092 \\
                          & TagFill\textsuperscript{D}  & \textbf{.416} & \textbf{.362} & \textbf{.353} & \textbf{.313} \\
  \bottomrule
  \end{tabular}
  \end{minipage}
\end{table*}

The $2\times 2$ encoding analysis reveals that encoding choice interacts strongly with method design.

\emph{Relative vs.\ absolute position.} For TagFill and IterEdit, absolute encoding is superior in most matched pairs (Table~\ref{tab:encoding_effect}a). Notably, this reverses the relative-over-absolute preference reported for autoregressive generation in the original \beat{} work~\cite{beat2025}: the ranking that holds under generation does not carry over to editing. The cause is \emph{cascading dependencies}: under relative encoding, modifying one pitch shifts all subsequent offsets within the beat, a bounded but compounding error that multi-pass methods amplify across refinement rounds. SeqTag, by contrast, applies all edits in a single pass and is immune to cascading---Scheme~B (Rel+Sep) achieves its \emph{highest} correction score, benefiting from relative encoding's translational invariance as a classification feature.

\emph{Separated vs.\ bundled tokens.} TagFill generally favors separated encoding (correction and completion), as its MLM-based inserter benefits from the richer per-token signal of two-token notes; editing is the exception, where bundled Scheme~D slightly leads. IterEdit's preference is task-dependent: separated for correction, bundled for editing, where shorter sequences reduce iterative passes. SeqTag is largely insensitive to this axis.

These two axes interact most dramatically at Scheme~B (Rel+Sep), where TagFill and IterEdit collapse below 0.10 while SeqTag achieves its highest score. The encoding$\times$method interaction is statistically significant (Appendix~\ref{app:significance}), confirming that encoding choice is a significant design lever alongside method architecture.

\subsection{External Comparison}\label{sec:boundary}

We compare TagFill against the diffusion model Polyffusion~\cite{min2023diffusion} on segment completion over its home dataset POP909 (4/4 pieces; song-level split: 161 train / 20 test songs, 204 eight-bar segments): given the melody and flanking accompaniment, fill four interior bars. The setup is conservative for us---Polyffusion's official checkpoint was trained on \emph{all} of POP909 including our test songs, whereas TagFill fine-tunes its inserter only on the training split---and both receive equal information (no ground-truth chords). Beyond reconstruction precision we report Yang--Lerch distribution-overlap metrics~\cite{yang2020evaluation} for pitch ($D_P$), duration ($D_D$), and inter-onset interval (IOI), plus chord-tone accuracy (CTA). TagFill leads reconstruction precision by ${\sim}10\times$ (bme) and ${\sim}4.6\times$ (nf1) and is better on $D_P$ and CTA, while Polyffusion produces more natural onset timing (Table~\ref{tab:polyffusion}): editing precisely reconstructs target content, diffusion generates distributionally natural timing. Given ground-truth chords, Polyffusion's CTA rises to .725, above the GT level of .681.

\begin{table}[t]
\caption{Segment completion on POP909 (204 segments). Poly.+chord additionally receives ground-truth chords.}
\label{tab:polyffusion}
\centering
\setlength{\tabcolsep}{3.2pt}
\begin{tabular}{@{}lccccccc@{}}
\toprule
\textbf{Method} & bme$\uparrow$ & nf1$\uparrow$ & $D_P\uparrow$ & $D_D\uparrow$ & IOI$\uparrow$ & CTA$\uparrow$ & ms$\downarrow$ \\
\midrule
TagFill-A & .083 & .254 & .907 & .860 & .680 & \textbf{.604} & 35 \\
TagFill-D & \textbf{.087} & \textbf{.279} & \textbf{.913} & \textbf{.924} & .805 & .576 & 35 \\
Polyffusion & .009 & .060 & .856 & .914 & \textbf{.867} & .427 & 1747 \\
\midrule
Poly.+chord & .013 & .084 & .889 & .930 & .945 & .725 & 1747 \\
\bottomrule
\end{tabular}
\end{table}

\subsection{Multi-Instrument Generalization}\label{sec:multi_instrument}

We construct a multi-track dataset from the Lakh MIDI Dataset~\cite{raffel2016lmd} (108K pieces, 105 General MIDI programs, 2--10 tracks per piece). Each piece is encoded using the Beat encoding's standard multi-track interleaving mechanism, where beats from all voices are interleaved at the beat level, and evaluated under Scheme~C. The framework transfers across all three tasks; full per-task metrics are given in Appendix~\ref{app:multi_track}.

Due to distributional differences between multi-track ensemble data and the dual-voice piano data used for training, evaluation metrics exhibit systematic shifts: beat exact match is lower (.400$\to$.258 on correction), reflecting the denser merged voices and greater harmonic complexity of ensemble arrangements, while note-level metrics are higher (nf1 .531$\to$.797), indicating that musical content is still captured under the shift. These results preliminarily show that the framework transfers to multi-instrument settings without architectural modification; a comprehensive evaluation across many heterogeneous tracks and instrument roles remains future work. We also tested a track-aware attention bias variant, which \emph{decreased} performance (Appendix~\ref{app:ablation}), suggesting shared attention already captures inter-track relationships without explicit structural bias.

\subsection{Discussion}\label{sec:discussion}

\textbf{Where editing yields to generation.}
Two experiments locate this boundary. Coarsely, at L4 (near-complete rewriting) all editing methods fall to 0.045--0.090 beat exact match (Table~\ref{tab:encoding_effect}b), and when accompaniment is removed entirely they collapse to 0.000. Finely, we sweep the gap width in the POP909 completion setting of \S\ref{sec:boundary}: keeping the melody and flanking accompaniment, we mask $N$ interior bars ($N{=}4$ to $16$; $N{=}16$ leaves no source accompaniment at all) and compute metrics only inside the gap. Degradation is two-phase. Reconstruction precision decreases monotonically---bme/nf1 fall from .100/.349 at $N{=}4$ to .051/.212 at $N{=}14$ and .041/.095 at $N{=}16$, with paired Wilcoxon tests (Holm-corrected) significant for both $N{=}8{\to}14$ and $N{=}14{\to}16$. Distributional quality, in contrast, holds whenever \emph{any} source context remains ($D_P$ at .88--.93 for all $N{\le}14$), then collapses from .878 to .623 at $N{=}16$---exactly when the last source content disappears. The paradigm's core advantage, preserving correct context, ceases to apply once no source survives: with no draft to edit, the task reverts to generation.

\textbf{Encoding--method interaction effects.}
Encoding choice (\S\ref{sec:encoding_effect}) also affects generative baselines: Diffusion exhibits extreme scheme dependence (D: 0.390 vs.\ A: 0.093 on correction), and AR-Prompt scores at most 0.15 regardless of scheme. As a practical guideline, relative-separated encoding (Scheme~B) is viable for single-pass classification, but should be avoided for iterative or generative pipelines where cascading errors compound.

\textbf{Edit localization as the AR bottleneck.}
We evaluate four AR (LLaMA) strategies with increasing positional guidance (Appendix~\ref{app:llama}). Performance scales monotonically with guidance: on editing, beat exact match rises from 0.095 (Prompt, no guidance) through 0.278 (Detect, SeqTag's error detector) to 0.499 (Selective, oracle positions). Notably, Teacher and Prompt perform comparably, suggesting that knowing the melody matters less than knowing \emph{where} to edit. Even with oracle positions, AR-Selective remains far behind edit-based methods on correction (0.477 vs.\ 0.726), indicating that AR generation cannot replicate per-token editing precision.

\textbf{Pre-training and architecture.}
Ablations (Appendix~\ref{app:ablation}) identify two critical design choices. First, removing BERT pre-training degrades all tasks substantially (correction: 0.700$\to$0.195), whereas the S1-only variant without task-specific fine-tuning stays comparable (0.719 vs.\ 0.700), indicating pre-trained representations are the primary driver. Second, replacing the encoder-only architecture with an encoder-decoder causes performance to collapse, confirming that shared bidirectional attention over both edited and context tokens is essential.

\label{sec:efficiency}\textbf{Inference efficiency.}
Single-pass methods (SeqTag, TagFill, BERT-CMLM) run in 37--65\,ms regardless of task and IterEdit in 82--655\,ms, one to two orders of magnitude faster than AR (9.4--23.7\,s). Per-task latency, per-encoding analysis, and an architecture-agnostic FLOPs estimate are given in Appendix~\ref{app:efficiency}; the latter shows the speed gap is paradigmatic rather than implementational, since editing needs only 1--3 forward passes.

%% ============================================================
%% 5. RELATED WORK
%% ============================================================
\section{Related Work}\label{sec:related}

\subsection{Symbolic Music Generation and Representation}
Symbolic music generation is dominated by two paradigms. Autoregressive Transformers~\cite{huang2018music,payne2019musenet,lu2023musecoco,vonrutte2023figaro} treat symbolic music as a flat sequence of discrete tokens---encoding pitch, duration, velocity, and timing as interleaved events---generated left-to-right. Diffusion models~\cite{min2023diffusion,huang2024ruleguided} instead denoise grid-based representations where pitch and time form a two-dimensional matrix. Both paradigms are tightly coupled with their underlying tokenization, making music representation a critical design decision. REMI~\cite{huang2020remi} uses multi-token note events, CPWord~\cite{hsiao2021compound} groups attributes into composites, and \beat{}~\cite{beat2025} proposes beat-anchored tokenization with ternary patterns. Systematic evaluations~\cite{fradet2023miditok,fradet2023impact,zhang2023symrep} confirm that tokenization choices significantly affect generation quality, as does work on compact multi-track encodings~\cite{dong2023mmt}.

\subsection{Edit-Based Generation}
Edit-based methods predict per-position operations, structurally preserving unedited content. In natural language processing, GECToR~\cite{omelianchuk2020gector} frames grammatical error correction as token-level tagging, LaserTagger~\cite{malmi2019lasertagger} combines KEEP/DELETE decisions with insertion phrases, the Levenshtein Transformer~\cite{gu2019levenshtein} performs iterative deletion and insertion, and Felix~\cite{mallinson2020felix} decomposes rewriting into tagging followed by filling; extensions include parallel iterative editing~\cite{awasthi2019parallel} and instruction-tuned models spanning diverse edit types~\cite{raheja2023coedit}. The same principle applies beyond text, to code bug fixing~\cite{dinella2020hoppity} and molecular optimization~\cite{jin2019learning}. Transferring it to music, however, is not immediate: text places one token per position, whereas a single beat carries several simultaneous pitches; under relative pitch encoding an edit to one note shifts its neighbour's position token; and notes held across beat boundaries span several token groups. The \tagshift{} operation (\S\ref{sec:error_correction}) and per-token editable flags (\S\ref{sec:task3}) are adaptations to precisely these properties.

\subsection{Music Editing}
Existing approaches to music editing are predominantly generative. Recent surveys cover both training-based and training-free audio editing with codec language models, diffusion, and flow matching~\cite{pan2026audioediting}. Music inpainting~\cite{pati2019inpainting} fills masked regions via conditional generation, with extensions to variable-length infilling~\cite{chang2021variable}, interactive refinement~\cite{hadjeres2021piano}, and multi-level control~\cite{guo2022musiac}. In the diffusion paradigm, SDEdit~\cite{meng2022sdedit} performs noise-then-denoise editing, GETMusic~\cite{lv2023getmusic} adopts discrete diffusion with a unified track-conditional representation, and cascaded diffusion models~\cite{wang2024wholesong} decompose whole-song synthesis into hierarchical inpainting stages. Text-guided methods manipulate diffusion latent spaces via textual instructions~\cite{zhang2024musicmagus} or edit audio with language-guided latent diffusion~\cite{wang2023audit}. For symbolic music, Not that Groove~\cite{zhang2025notthatgroove} uses zero-shot LLM prompting over a textual drumroll notation to edit single-bar drum grooves, with instruction adherence checked by symbolic unit tests. These methods generate revised content conditioned on context or instructions; \beatedit{} learns explicit per-position edit operations over a beat-anchored symbolic encoding.

%% ============================================================
%% 6. CONCLUSION
%% ============================================================
\section{Conclusion}\label{sec:conclusion}

This work demonstrates that edit-based methods are effective for symbolic music generation, achieving higher precision and perceptual quality than autoregressive and diffusion baselines across all three tasks, with single-pass inference under 100\,ms. Our cross-factorial analysis reveals encoding design as a critical, previously overlooked performance lever: effective editing requires atomic edit units, trivial source--target alignment, and bounded edit locality, and encoding choice can rival method choice. The three mechanisms specialize complementarily along the edit-density axis---sequence tagging for sparse corrections, iterative refinement for medium-density editing, tag-then-fill for segment completion---while performance degrades sharply under near-complete rewriting, delineating a principled boundary where editing yields to generation.

Our experiments rely on synthetically perturbed data (\S\ref{sec:setup}). While this allows precise control over task difficulty and perturbation ratios, it does not fully capture the diversity of real-world music editing, where drafts are revised, arrangements varied, and styles adapted. We hope this work motivates the community to build large-scale paired datasets of genuine compositional edits.

%% ============================================================
%% ACKNOWLEDGMENTS
%% ============================================================
\begin{acks}
This work was supported in part by the ZJ program of Guangdong (2023QN10X455), the Fundamental Research Funds for the Central Universities (2025ZYGXZR053), the GJYC program of Guangzhou (2024D01J0081), the National Natural Science Foundation of China (62401377), and the Yangtze River Delta Science and Technology Innovation Community Joint Research Project (2024CSJGG1100).
\end{acks}

%% ============================================================
%% BIBLIOGRAPHY
%% ============================================================
\bibliographystyle{ACM-Reference-Format}
\bibliography{references}

\newpage
\appendix

\section{Encoding Technical Specification}\label{app:encoding_spec}

\subsection{Complete Token Vocabulary}

\begin{table}[ht]
\caption{Complete token ID allocation for separated encoding (Schemes A, B). Under bundled encoding (Schemes C, D), position and pattern tokens merge into composite tokens (IDs 0--7{,}128), with separate control tokens.}
\label{tab:full_vocab}
\centering
\small
\begin{tabular}{@{}lrl@{}}
\toprule
\textbf{Token Type} & \textbf{ID Range} & \textbf{Description} \\
\midrule
Pattern values & 0--80 & Ternary rhythm patterns ($3^4 = 81$) \\
Position markers & 81--168 & $81 + \text{pitch\_index}$ (max 87) \\
EMPTY\_MARKER & 169 & Empty beat (single token) \\
BAR & 170 & Bar separator \\
EOS & 171 & End of sequence \\
BOS & 172 & Beginning of sequence \\
PAD & 173 & Padding token \\
TIME\_SIG & 174--178 & 5 time signatures (4/4, 3/4, 2/4, 6/8, 2/2) \\
BPM & 179--182 & 4 tempo categories \\
T0\_START & 183 & Right-hand voice marker \\
T1\_START & 184 & Left-hand voice marker \\
MASK & 185 & MLM pre-training only \\
\bottomrule
\end{tabular}
\end{table}

\subsection{Ternary Pattern Semantics}

The within-beat state vector $\mathbf{s}_p \in \{0,1,2\}^\tau$ of pitch $p$ (0~=~silent, 1~=~onset, 2~=~sustain) is mapped to a single pattern token by base-3 conversion:
\begin{equation}\label{eq:patch}
  s_p = \sum_{t=0}^{\tau-1} \mathbf{s}_p[t] \cdot 3^{\tau-1-t} \in \{0,\dots,3^\tau-1\}.
\end{equation}
For $\tau = 4$ this yields $81$ values; e.g.\ a quarter note $(1,2,2,2) \mapsto 1{\cdot}27 + 2{\cdot}9 + 2{\cdot}3 + 2 = 53$ and two eighth notes $(1,2,1,2) \mapsto 50$. Selected pattern values for $\tau = 4$:

\begin{center}
\small
\begin{tabular}{@{}ccl@{}}
\toprule
\textbf{Ternary} $\mathbf{s}_p$ & $\boldsymbol{s_p}$ & \textbf{Musical Meaning} \\
\midrule
$(1,2,2,2)$ & 53 & Quarter note (onset + sustain) \\
$(1,2,1,2)$ & 50 & Two eighth notes \\
$(1,0,0,0)$ & 27 & Sixteenth note (staccato) \\
$(0,0,1,2)$ & 5 & Note on beat subdivision 3 \\
$(2,2,2,2)$ & 80 & Sustained from previous beat \\
$(0,0,0,0)$ & 0 & Silent (not encoded) \\
\bottomrule
\end{tabular}
\end{center}

\subsection{Encoding and Decoding Algorithms}

\begin{algorithm}[ht]
\caption{\beat{} Encoding (Single-Track)}
\label{alg:beat_encoding}
\begin{algorithmic}[1]
\REQUIRE piano-roll $X \in \{0,1,2\}^{P \times T}$, steps per beat $\tau$, beats per bar $n$
\ENSURE Token sequence $\mathbf{E}$
\STATE $N \gets \lceil T / \tau \rceil$ \COMMENT{Number of beats}
\STATE $\mathbf{E} \gets [\,]$
\FOR{$i = 1$ to $N$}
    \IF{$(i-1) \bmod n = 0$}
        \STATE Append \texttt{BAR} to $\mathbf{E}$
    \ENDIF
    \STATE Append \texttt{BEAT} to $\mathbf{E}$
    \STATE $B^{(i)} \gets X[:, (i{-}1)\tau : i\tau]$ \COMMENT{Beat segment}
    \STATE $\mathcal{A} \gets \{p : \exists\, t,\; B^{(i)}[p, t] \neq 0\}$ \COMMENT{Active pitches}
    \IF{$\mathcal{A} = \emptyset$}
        \STATE Append \texttt{EMPTY\_MARKER} to $\mathbf{E}$ and continue
    \ENDIF
    \STATE Sort $\mathcal{A}$ ascending: $p_1 < p_2 < \cdots < p_M$
    \FOR{$j = 1$ to $M$}
        \STATE $\mathbf{s}_{p_j} \gets B^{(i)}[p_j, :]$ \COMMENT{State vector}
        \STATE $s_{p_j} \gets \textsc{Base3ToInt}(\mathbf{s}_{p_j})$ \COMMENT{Pattern token}
        \IF{$j = 1$}
            \STATE $d_1 \gets p_1$ \COMMENT{Absolute pitch}
        \ELSE
            \STATE $d_j \gets p_j - p_{j-1}$ \COMMENT{Relative interval}
        \ENDIF
        \STATE Append $(\texttt{PIT}\,d_j,\; \texttt{PAT}\,s_{p_j})$ to $\mathbf{E}$
    \ENDFOR
\ENDFOR
\STATE \textbf{return} $\mathbf{E}$
\end{algorithmic}
\end{algorithm}

\begin{algorithm}[ht]
\caption{\beat{} Decoding (Single-Track)}
\label{alg:beat_decoding}
\begin{algorithmic}[1]
\REQUIRE Token sequence $\mathbf{E}$, steps per beat $\tau$
\ENSURE piano-roll $X \in \{0,1,2\}^{P \times T}$
\STATE Initialize $X \gets \mathbf{0}^{P \times T}$
\STATE $i \gets 0$, \quad $p_{\text{prev}} \gets \textsc{None}$
\FOR{each token in $\mathbf{E}$}
    \IF{token is \texttt{BAR}}
        \STATE Continue
    \ELSIF{token is \texttt{BEAT}}
        \STATE $i \gets i + 1$, \quad $p_{\text{prev}} \gets \textsc{None}$
    \ELSIF{token is \texttt{EMPTY\_MARKER}}
        \STATE Continue
    \ELSIF{token is $(\texttt{PIT}\,d,\; \texttt{PAT}\,s)$}
        \IF{$p_{\text{prev}} = \textsc{None}$}
            \STATE $p \gets d$ \COMMENT{Absolute pitch (first in beat)}
        \ELSE
            \STATE $p \gets p_{\text{prev}} + d$ \COMMENT{Relative pitch}
        \ENDIF
        \IF{$0 \leq p < P$}
            \STATE $X[p, (i{-}1)\tau : i\tau] \gets \textsc{IntToBase3}(s, \tau)$
        \ENDIF
        \STATE $p_{\text{prev}} \gets p$
    \ENDIF
\ENDFOR
\STATE \textbf{return} $X$
\end{algorithmic}
\end{algorithm}

\section{Edit Label Space}\label{app:labels}

\begin{table}[ht]
\caption{Complete 350-label edit space for separated encoding (Schemes A, B). Under bundled encoding, $\tagreplace{}$ and $\tagappend{}$ target 7{,}128 music tokens and \tagshift{} is absent, yielding 14{,}258 labels.}
\label{tab:label_functions}
\centering
\begin{tabular}{@{}llrl@{}}
\toprule
\textbf{Operation} & \textbf{Label ID} & \textbf{Count} & \textbf{Computation} \\
\midrule
$\tagkeep{}$ & 0 & 1 & --- \\
$\tagdelete{}$ & 1 & 1 & --- \\
$\tagreplace{}_v$ & $2 + v$ & 169 & $v \in [0, 168]$ \\
$\tagappend{}_v$ & $171 + v$ & 169 & $v \in [0, 168]$ \\
$\tagshift{}_{+n}$ & $339 + n$ & 5 & $n \in [1, 5]$ \\
$\tagshift{}_{-n}$ & $344 + |n|$ & 5 & $n \in [-5, -1]$ \\
\bottomrule
\end{tabular}
\end{table}

\section{Music BERT Pre-Training Details}\label{app:bert}
\begin{table}[ht]
\caption{Music BERT architecture and pre-training configuration.}
\label{tab:bert_pretrain}
\centering
\small
\setlength{\tabcolsep}{4pt}
\begin{tabular}{@{}ll@{}}
\toprule
\textbf{Component} & \textbf{Configuration} \\
\midrule
Layers / Heads / Hidden / FFN & 8 / 8 / 512 / 2,048 \\
Parameters & $\sim$26.6M \\
Vocab size & 185 (184 + \texttt{MASK}) \\
Max position embeddings & 2,048 \\
\midrule
Training data & 192,788 pieces (154K / 19.3K / 19.3K split) \\
Masking & 15\% of music tokens \\
Batch size / Epochs & 256 effective / 30 \\
Optimizer / Schedule & AdamW, lr $10^{-4}$, cosine + 10\% warmup \\
\bottomrule
\end{tabular}
\end{table}

The Music BERT encoder uses the architecture specified in Table~\ref{tab:bert_pretrain}. Pre-training uses standard MLM: 15\% of music tokens are selected, of which 80\% are replaced with \texttt{[MASK]}, 10\% with a random music token, and 10\% left unchanged. Control tokens (IDs $\geq 169$) are never masked. Data augmentation applies random transposition ($\pm 5$ semitones) with 70\% probability; sequences exceeding 2,048 tokens are randomly truncated.

\textbf{Training dynamics.} Best validation losses (across all epochs) vary across encoding schemes: Scheme~B achieves the lowest loss (0.224, perplexity 1.25), followed by A~(0.282, ppl~1.33), C~(0.449, ppl~1.57), and D~(0.502, ppl~1.65). The lower perplexity of separated schemes reflects their smaller vocabulary (185 vs.\ 7{,}145). Training times range from 4.6h (Scheme~D) to 27.3h (Scheme~B) on 2~GPUs (24GB VRAM each).

\section{Implementation Details}\label{app:implementation}

\subsection{Head and Loss Formulations}\label{app:formulations}

Let $\mathbf{h}_i \in \mathbb{R}^d$ denote the shared encoder's hidden representation at position $i$. All prediction heads are single linear layers over $\mathbf{h}_i$.

\textbf{SeqTag} (\S3.2 of the main paper). The binary error-detection head and the multi-class tag head are
\begin{equation}\label{eq:seqtag_heads}
  \hat{e}_i = \sigma(\mathbf{W}_{\text{det}}\,\mathbf{h}_i + \mathbf{b}_{\text{det}}), \qquad
  \hat{t}_i = \mathrm{softmax}(\mathbf{W}_{\text{tag}}\,\mathbf{h}_i + \mathbf{b}_{\text{tag}}),
\end{equation}
trained jointly with $\mathcal{L} = \mathcal{L}_{\text{tag}} + \lambda \cdot \mathcal{L}_{\text{detect}}$, where $\mathcal{L}_{\text{tag}}$ is weighted cross-entropy with down-weighted \tagkeep{} and $\mathcal{L}_{\text{detect}}$ is binary cross-entropy. At inference $\hat{t}_i$ is overridden to \tagkeep{} wherever $\hat{e}_i < \theta$.

\textbf{IterEdit} (\S3.3). The deletion, insertion, and token heads are
\begin{equation}\label{eq:levt_heads}
  \begin{aligned}
    \hat{d}_i &= \sigma(\mathbf{W}_{\text{del}}\,\mathbf{h}_i), \\
    \hat{n}_j &= \mathrm{softmax}(\mathbf{W}_{\text{ins}}\,[\mathbf{h}_j; \mathbf{h}_{j+1}]), \\
    \hat{x}_i &= \mathrm{softmax}(\mathbf{W}_{\text{tok}}\,\mathbf{h}_i).
  \end{aligned}
\end{equation}
where $\hat{n}_j$ predicts the number of placeholders (0--20) to insert in the gap $j$ between adjacent tokens. The joint loss $\mathcal{L}_{\text{iter}} = \mathcal{L}_{\text{del}} + \mathcal{L}_{\text{ins}} + \mathcal{L}_{\text{tok}}$ sums three cross-entropy terms; token prediction uses label smoothing ($\epsilon{=}0.1$).

\textbf{TagFill} (\S3.4). The tagger is trained with Focal Loss
\begin{equation}\label{eq:focal}
  \mathcal{L}_{\text{tagger}} = -\sum_{i} (1 - \hat{p}_{i,y_i})^\gamma \log \hat{p}_{i,y_i}, \qquad \gamma = 2.0,
\end{equation}
and the inserter commits $\hat{x}_i = \arg\max_v P_{\text{ins}}(v \mid \tilde{X})$ for $i \in \mathcal{M}_t$, the confidence-ranked subset of masks accepted at pass $t$.

\subsection{Configurations}

\begin{table}[ht]
\caption{SeqTag configuration.}
\label{tab:config_gector}
\centering
\small
\begin{tabular}{@{}ll@{}}
\toprule
\textbf{Component} & \textbf{Configuration} \\
\midrule
\multicolumn{2}{@{}l}{\textit{Heads}} \\
\quad Error detector & Linear(512, 2) \\
\quad Tag predictor & Linear(512, 350 / 14{,}258) \\
\quad Dropout & 0.1 \\
\midrule
\multicolumn{2}{@{}l}{\textit{Training}} \\
\quad Cold start & 2 epochs, freeze BERT (lr $10^{-3}$) \\
\quad Full fine-tune & 18 epochs, patience 3 \\
\quad Fine-tune lr (BERT / heads) & $10^{-5}$ / $10^{-4}$ \\
\quad Batch size / Precision & 64 effective / fp16 \\
\quad \tagkeep{} weight / $\lambda_{\text{detect}}$ & 0.15 / 0.5 \\
\midrule
\multicolumn{2}{@{}l}{\textit{Inference}} \\
\quad Max iterations & 3 \\
\quad KEEP bias $\beta$ / Error threshold $\theta$ & 0.3 / 0.5 \\
\bottomrule
\end{tabular}
\end{table}

\begin{table}[ht]
\caption{TagFill (tag-then-fill) configuration.}
\label{tab:config_felix}
\centering
\small
\begin{tabular}{@{}ll@{}}
\toprule
\textbf{Component} & \textbf{Configuration} \\
\midrule
\multicolumn{2}{@{}l}{\textit{Tagger}} \\
\quad Architecture & 8-layer Transformer, 512 hidden \\
\quad Output head & Linear(512, 11) \\
\quad Loss & Focal Loss ($\gamma = 2.0$) \\
\midrule
\multicolumn{2}{@{}l}{\textit{Inserter}} \\
\quad Architecture & 8-layer Transformer, 512 hidden \\
\quad Output head & Linear(512, $V$) \\
\quad Label smoothing & 0.1 \\
\quad MaskGIT passes $T$ & 2 \\
\midrule
\multicolumn{2}{@{}l}{\textit{Training (both stages)}} \\
\quad Optimizer / Schedule & AdamW, lr $10^{-4}$, cosine + 10\% warmup \\
\quad Weight decay / Epochs & 0.01 / 30 \\
\quad Batch size & 96 effective (32 $\times$ 3 accum.) \\
\quad Perturbation levels & L1:30\%, L2:30\%, L3:25\%, L4:15\% \\
\bottomrule
\end{tabular}
\end{table}

\begin{table}[ht]
\caption{IterEdit (iterative deletion-insertion) configuration.}
\label{tab:config_levt}
\centering
\small
\begin{tabular}{@{}ll@{}}
\toprule
\textbf{Component} & \textbf{Configuration} \\
\midrule
\multicolumn{2}{@{}l}{\textit{Architecture}} \\
\quad Encoder & 8-layer Transformer, 512 hidden, 8 heads \\
\quad Activation / Norm & GELU / Pre-norm \\
\quad Deletion head & Linear(512, 2) \\
\quad Insertion head & Linear(1024, 21) \\
\quad Token head & Linear(512, $V$) \\
\midrule
\multicolumn{2}{@{}l}{\textit{Training}} \\
\quad Optimizer / Schedule & AdamW, lr $3 \times 10^{-4}$, cosine + 10\% warmup \\
\quad Weight decay / Epochs & 0.01 / 30 \\
\quad Batch size & 64 effective (32 $\times$ 2 accum.) \\
\quad Loss & $\mathcal{L}_{\text{del}} + \mathcal{L}_{\text{ins}} + \mathcal{L}_{\text{tok}}$ \\
\quad Token label smoothing & 0.1 \\
\midrule
\multicolumn{2}{@{}l}{\textit{Inference}} \\
\quad Max iterations & 10 (completion) / 3 (editing) \\
\quad Deletion threshold & 0.5 \\
\bottomrule
\end{tabular}
\end{table}

\section{SeqTag Inference Algorithm}\label{app:seqtag_alg}

\begin{algorithm}[h]
\caption{Iterative music error correction (SeqTag)}
\label{alg:inference}
\begin{algorithmic}[1]
\REQUIRE Encoded tokens $X$, model $\mathcal{M}$, max rounds $R_{\max}=3$
\FOR{$r = 1$ to $R_{\max}$}
  \STATE $d \leftarrow \mathcal{M}_{\text{detect}}(X)$ \COMMENT{binary error detection}
  \STATE $t \leftarrow \mathcal{M}_{\text{tag}}(X)$ \COMMENT{edit tag prediction}
  \STATE Gate: $t_i \leftarrow \tagkeep{}$ wherever $d_i < 0.5$ \COMMENT{conservative KEEP bias}
  \IF{all $t_i = \tagkeep{}$}
    \STATE \textbf{break}
  \ENDIF
  \STATE $X_{\mathrm{prev}} \leftarrow X$
  \STATE Apply labels: $X \leftarrow \textsc{ApplyEdits}(X, t)$
  \STATE $X \leftarrow \textsc{Decode-Filter-Reencode}(X)$ \COMMENT{validity check}
  \IF{$X = X_{\mathrm{prev}}$}
    \STATE \textbf{break}
  \ENDIF
\ENDFOR
\RETURN $X$
\end{algorithmic}
\end{algorithm}

The \textsc{Decode-Filter-Reencode} step operates beat by beat: it decodes each beat's tokens into (pitch, pattern) notes, discards invalid ones (out-of-range pitches or pattern values, duplicate pitches within a beat), and re-encodes the remainder in ascending pitch order, which also repairs any ordering violations. A beat that fails to parse is left unchanged, and beats outside the edited region reassemble verbatim, so the cycle guarantees a well-formed input for the next iteration without perturbing untouched content.

\section{End-to-End Error Correction Walkthrough}\label{app:walkthrough}

We trace a correction involving two error types---a pitch shift and a missing note---through the full pipeline under Scheme~B (relative, separated). Consider a beat whose target is C4--E4--G4 as quarter notes (pattern token \texttt{53}, Appendix~\ref{app:encoding_spec}). The input has C4 shifted to D4 ($+2$ semitones) and G4 missing.

\textbf{Step 1: Encoding.} The target and corrupted input are:
\begin{center}
\begin{tabular}{@{}ll@{}}
\toprule
\textbf{Target} & \texttt{[120][53]\ [85][53]\ [84][53]} \\
& \footnotesize (C4, E4, G4 as quarter notes) \\
\textbf{Input} & \texttt{[122][53]\ [83][53]} \\
& \footnotesize (D4, E4 only---G4 missing) \\
\bottomrule
\end{tabular}
\end{center}
D4 yields position $81+41=122$; the relative distance from D4 to E4 is $2$ (token \texttt{83}).

\textbf{Step 2: Iteration 1---pitch correction via \tagshift{}.} The model detects the pitch error and predicts:
\begin{center}
\begin{tabular}{@{}lcccc@{}}
\toprule
\textbf{Token} & \texttt{[122]} & \texttt{[53]} & \texttt{[83]} & \texttt{[53]} \\
\midrule
\textbf{Label} & $\tagshift{}_{-2}$ & \tagkeep{} & $\tagshift{}_{+2}$ & $\tagappend{}_{84}$ \\
\textbf{Result} & \texttt{[120]} & \texttt{[53]} & \texttt{[85]} & \texttt{[53]}\ \texttt{[84]} \\
\bottomrule
\end{tabular}
\end{center}
$\tagshift{}_{-2}$ corrects D4$\to$C4; $\tagshift{}_{+2}$ fixes the cascading interval to E4; $\tagappend{}_{84}$ inserts a position token for the missing G4 ($81+3=84$, relative to E4). Since \tagappend{} attaches only \emph{one} token per anchor, the pattern token for G4 cannot be produced within the same round.

\textbf{Step 3: Post-processing and convergence.} The \textsc{Decode-Filter-Reencode} step (Appendix~\ref{app:seqtag_alg}) parses the beat into complete (position, pattern) pairs; the trailing token \texttt{[84]} lacks a pattern token, does not form a valid note, and is discarded during re-encoding, yielding \texttt{[120][53]\ [85][53]}. Both pitch errors are therefore fixed within a single iteration, while the missing G4 is \emph{not} restored: even if a later iteration proposes the same \tagappend{}, post-processing removes the incomplete insertion again and the no-progress check terminates inference. This reflects a structural limitation of separated encodings---restoring a deleted note requires two coordinated tokens, but validity filtering only admits complete pairs---and is consistent with note deletion being the hardest error type for Schemes~A and~B (Figure~\ref{fig:error_type}). Under bundled encoding the same repair is a single edit: in Scheme~C the missing G4 corresponds to one composite token ($3 \times 81 + 53 = 296$), so $\tagappend{}_{296}$ completes the insertion within one iteration and passes post-processing unchanged. This walkthrough demonstrates that \tagshift{} and \tagappend{} cooperate within a single iteration, and that \textsc{Decode-Filter-Reencode} acts as a structural safety net: every intermediate sequence remains well-formed and decodable, at the cost of discarding insertions left incomplete at the end of an iteration.

\section{Synthetic Perturbation Details}\label{app:perturbation}

\begin{table}[ht]
\caption{Perturbation types for synthetic error data. Per-beat, mutually exclusive.}
\label{tab:perturbation}
\centering
\small
\setlength{\tabcolsep}{4pt}
\begin{tabular}{@{}llcl@{}}
\toprule
\textbf{Type} & \textbf{Operation} & \textbf{Prob.} & \textbf{Constraint} \\
\midrule
Pitch shift & $p_j \leftarrow p_j + \text{Uniform}(\pm 1\text{--}3)$ & 0.10 & Valid range \\
Rhythm change & $s_j \leftarrow \text{Random}([0, 80] \setminus \{s_j\})$ & 0.05 & --- \\
Note deletion & Remove one $(p_j, s_j)$ pair & 0.03 & Keep $\geq 1$ note \\
Note insertion & Add $(p_{\text{new}}, s_{\text{new}})$ & 0.02 & No duplicates \\
No change & Identity & 0.80 & --- \\
\bottomrule
\end{tabular}
\end{table}

\section{IterEdit Inference Algorithm}\label{app:iteredit_alg}

\begin{algorithm}[h]
\caption{Iterative editing inference (IterEdit)}
\label{alg:levt}
\begin{algorithmic}[1]
\REQUIRE Current tokens $X$, editable flags $E$, model $\mathcal{M}$, max rounds $R_{\max}=10$
\FOR{$r = 1$ to $R_{\max}$}
  \STATE \textit{// Step 1: Delete low-confidence tokens}
  \STATE $d \leftarrow \mathcal{M}_{\text{del}}(X)$; remove $X_i$ where $d_i{>}0.5$ \textbf{and} $E_i{=}\text{True}$
  \STATE \textit{// Step 2: Insert placeholders at editable gaps}
  \STATE $n \leftarrow \mathcal{M}_{\text{ins}}(X)$; insert $n_j$ \texttt{PLH} tokens at gap $j$ if gap is editable
  \STATE Mark all new \texttt{PLH} positions as editable
  \STATE \textit{// Step 3: Fill placeholders}
  \STATE $X_i \leftarrow \arg\max \mathcal{M}_{\text{tok}}(X)_i$ for each \texttt{PLH} position $i$
  \IF{no deletions, no insertions, and no fills}
    \STATE \textbf{break}
  \ENDIF
\ENDFOR
\RETURN $X$
\end{algorithmic}
\end{algorithm}

\section{Generative Baseline Details}\label{app:baselines}

The main text reports summary results for the two generative baselines; here we provide full per-scheme and per-configuration breakdowns that reveal encoding sensitivity patterns consistent with those observed for edit-based methods.

\subsection{Autoregressive LLM Baseline}\label{app:llama}

To quantify the gap between autoregressive generation and edit-based paradigms, we evaluate four LLaMA-based strategies (${\sim}$162M parameters, trained from scratch on the same data), ordered by the amount of positional guidance provided:

\begin{itemize}[nosep,leftmargin=*]
  \item \textbf{AR-Prompt}: takes the first 2~bars of the source as prompt and autoregressively generates the rest; the generated melody is replaced with the source melody post hoc.
  \item \textbf{AR-Teacher}: melody beats are fed via teacher-forcing with explicit track markers; \emph{all} accompaniment beats are regenerated.
  \item \textbf{AR-Detect}: a two-stage pipeline that uses SeqTag's error detection head to identify perturbed beats (any accompaniment token with $P(\text{error}) \geq 0.5$ flags the beat), then regenerates only those beats via LLaMA with melody teacher-forcing. This provides the fairest AR comparison, as it uses the same localization signal available to edit methods.
  \item \textbf{AR-Selective}: same teacher-forcing as Teacher, but only the accompaniment beats at oracle-known \texttt{changed\_beat\_indices} are regenerated; unperturbed beats retain the source content. This is the most favorable setting for AR.
\end{itemize}

\noindent\textbf{Implementation.} For Scheme~A, melody beats start with \texttt{Track0}~(183) and accompaniment with \texttt{Track1}~(184); Teacher/Selective explicitly insert~184 before sampling accompaniment content. For bundled schemes~(C/D), \texttt{SPLIT\_0}~(7129) and \texttt{SPLIT\_1}~(7130) serve as track markers. For Scheme~B (no explicit marker), beats alternate by position and terminate with \texttt{END\_MARKER}~(170). Prompt/Full post-processing parses the generated sequence into beat pairs and replaces melody beats with source melody.

\noindent\textbf{Fairness note.} Edit methods (SeqTag/TagFill/IterEdit) see the complete corrupted source and must autonomously identify which positions need modification. AR-Prompt and AR-Teacher do not know which beats are perturbed and regenerate all accompaniment. AR-Selective additionally receives oracle \texttt{changed\_beat\_indices}---information unavailable to edit methods. Results for Selective therefore represent an \emph{upper bound} on AR performance under ideal localization, not a fair head-to-head comparison.

\begin{table}[ht]
\caption{LLaMA baseline results ($n=200$, perturbed-only). Best per task--strategy in \textbf{bold}. $^\dagger$Selective receives oracle changed-beat positions.}
\label{tab:llama_results}
\centering
\small
\begin{tabular}{@{}llccccc@{}}
\toprule
\textbf{Strategy} & \textbf{Task} & \textbf{Sch.} & \textbf{beat}$\uparrow$ & \textbf{nf1}$\uparrow$ & \textbf{MPE}$\downarrow$ & \textbf{FMD}$\downarrow$ \\
\midrule
\multirow{3}{*}{AR-Select.$^\dagger$}
  & Corr.  & A & \textbf{.477} & \textbf{.637} & \textbf{8.6}  & \textbf{1.17} \\
  & Edit.  & A & \textbf{.499} & \textbf{.603} & \textbf{9.6}  & \textbf{0.85} \\
  & Compl. & A & \textbf{.397} & \textbf{.512} & \textbf{14.4} & \textbf{0.39} \\
\midrule
\multirow{2}{*}{AR-Detect}
  & Corr.  & A & .394 & .650 & 9.1  & 1.24 \\
  & Edit.  & A & .278 & .478 & 15.2 & 1.73 \\
\midrule
\multirow{3}{*}{AR-Teacher}
  & Corr.  & B & .062 & .238 & 20.0 & 5.74 \\
  & Edit.  & A & .101 & .211 & 25.6 & 7.07 \\
  & Compl. & C & .117 & .221 & 22.6 & 5.82 \\
\midrule
\multirow{3}{*}{AR-Prompt}
  & Corr.  & D & .093 & .242 & 16.3 & 11.2 \\
  & Edit.  & A & .095 & .211 & 25.8 & 3.89 \\
  & Compl. & A & .152 & .221 & 18.7 & 1.54 \\
\bottomrule
\end{tabular}
\end{table}

The results reveal a clear \emph{localization hierarchy}. Prompt and Teacher, which lack positional guidance, achieve at most 0.152 beat exact match. Detect, using SeqTag's error detector for automated localization, jumps to 0.394 on correction and 0.278 on editing---a substantial improvement but still below edit methods (SeqTag: 0.726, IterEdit: 0.480). Selective, given oracle positions, reaches 0.477--0.499 but still cannot match per-token precision on correction (0.477 vs.\ 0.726). Detect is inapplicable to completion, as the error detector cannot identify missing content.

The gap between Detect and edit methods is particularly informative: both use the same localization signal (SeqTag's detector), but edit methods apply targeted token-level operations while Detect regenerates entire beats via AR sampling. This confirms that the editing mechanism itself---not just localization---is the source of the performance advantage. Selective consistently favors Scheme~A, and its FMD values (0.39--1.17) are competitive with edit methods, suggesting that AR generation can produce distributionally natural output at the cost of requiring oracle positional guidance.

\subsection{Discrete Diffusion Baseline}\label{app:diffusion}

We evaluate D3PM~\cite{austin2021d3pm} discrete diffusion (${\sim}$34M parameters) with an SDEdit-style noise-then-denoise strategy at three noise ratios: $r{=}0.3$ (30 denoising steps), $r{=}0.5$ (50 steps), and $r{=}0.7$ (70 steps). All models are trained on the same data with all four encoding schemes ($n=200$, perturbed-only).

\begin{table*}[t]
\caption{Diffusion baseline results across noise ratios and encoding schemes ($n=200$, perturbed-only). Best value for each metric within each ratio--task in \textbf{bold}.}
\label{tab:diffusion_results}
\centering
\setlength{\tabcolsep}{4pt}
\begin{tabular}{@{}llcccccccccccc@{}}
\toprule
& & \multicolumn{4}{c}{\textbf{Error Correction}} & \multicolumn{4}{c}{\textbf{Accomp.\ Editing}} & \multicolumn{4}{c}{\textbf{Segment Completion}} \\
\cmidrule(lr){3-6}\cmidrule(lr){7-10}\cmidrule(lr){11-14}
\textbf{Ratio} & \textbf{Sch.} & beat & nf1 & FMD & fail & beat & nf1 & FMD & fail & beat & nf1 & FMD & fail \\
\midrule
\multirow{4}{*}{$r{=}0.3$}
  & A & .093 & .190 & 14.1 & .78 & .076 & .134 & 18.20 & .82 & .053 & .168 & \textbf{14.9} & .99 \\
  & B & .048 & .124 & 17.4 & .87 & .075 & .088 & 22.08 & .80 & .050 & .119 & 19.6 & .97 \\
  & C & .340 & .515 & 2.03 & .07 & \textbf{.200} & .372 & 2.75 & \textbf{.22} & \textbf{.080} & \textbf{.228} & 52.7 & \textbf{.65} \\
  & D & \textbf{.390} & \textbf{.602} & \textbf{0.87} & \textbf{.04} & .198 & \textbf{.411} & \textbf{1.72} & .23 & \textbf{.080} & \textbf{.228} & 37.5 & \textbf{.65} \\
\midrule
\multirow{4}{*}{$r{=}0.5$}
  & A & .029 & .013 & 217 & .98 & .081 & .010 & 221 & .98 & .028 & .000 & 222 & .99 \\
  & B & .018 & .005 & 171 & .99 & .070 & .010 & 172 & .95 & .005 & .000 & 173 & 1.0 \\
  & C & .153 & .294 & 3.90 & .37 & .095 & .186 & 5.91 & .50 & .071 & \textbf{.002} & \textbf{4.79} & \textbf{.68} \\
  & D & \textbf{.188} & \textbf{.358} & \textbf{3.32} & \textbf{.29} & \textbf{.105} & \textbf{.230} & \textbf{5.43} & \textbf{.43} & \textbf{.074} & \textbf{.002} & 4.84 & .70 \\
\midrule
\multirow{4}{*}{$r{=}0.7$}
  & A & .020 & .002 & 453 & 1.0 & \textbf{.101} & .000 & 454 & 1.0 & .013 & .000 & 455 & 1.0 \\
  & B & .015 & .000 & 355 & .99 & .085 & .000 & 353 & .93 & .003 & .000 & 359 & 1.0 \\
  & C & .057 & .133 & \textbf{7.92} & .76 & .046 & .085 & \textbf{10.9} & .79 & .060 & .002 & \textbf{10.2} & .79 \\
  & D & \textbf{.067} & \textbf{.191} & 10.2 & \textbf{.67} & .063 & \textbf{.113} & 14.5 & \textbf{.71} & \textbf{.061} & \textbf{.003} & 11.1 & \textbf{.78} \\
\bottomrule
\end{tabular}
\end{table*}

Three findings are notable.
First, \emph{performance degrades sharply with noise ratio}: $r{=}0.3$ is the only competitive setting; $r{=}0.5$ halves beat exact match, and $r{=}0.7$ collapses to near-zero on separated encodings (FMD~$>$200, fail rate~$>$93\%). This confirms that SDEdit requires minimal noise injection for editing---heavy noise destroys the structural information that denoising cannot recover.
Second, \emph{encoding dependence is extreme and mirrors edit methods}: bundled schemes (C/D) outperform separated schemes (A/B) by $4$--$8\times$ on correction (r03: Scheme~D 0.390 vs.\ A 0.093). Separated encoding distributes each note across two tokens; the diffusion process must independently recover both tokens in a consistent pair, a coordination problem that bundled encoding avoids. This parallel to edit methods' encoding sensitivity provides independent evidence that encoding choice is a universal design lever, not method-specific.
Third, \emph{completion exposes diffusion's fundamental limitation}: Diffusion~r03 maintains non-trivial beat exact match on completion (0.080) but low nf1 (0.228), and at higher noise ratios ($r{=}0.5$) nf1 collapses to 0.002, indicating that the denoising process recovers rhythmic structure but struggles with pitch content. TagFill's nf1 of 0.557 on the same task confirms that explicit generation mechanisms are required for content synthesis.

\section{Cross-Encoding Pre-Training Analysis}\label{app:cross_encoding}

Table~3 (main paper) reports a \emph{recovery} metric defined as $(\text{nf1} - \text{nf1}_{\text{no-edit}}) / (1 - \text{nf1}_{\text{no-edit}})$, measuring the fraction of the no-edit error gap closed by the method. \beat{} achieves ${\sim}5\times$ higher recovery than alternative encodings. Here we trace this gap to its root cause: encoding structure determines pre-training quality, which cascades into downstream editing performance.

\subsection{MLM Pre-Training Convergence}

\begin{table}[ht]
\caption{MLM pre-training convergence across encodings. Values are eval perplexity ($\downarrow$). PPL values are not directly comparable across encodings with different vocabulary structures (see text).}
\label{tab:mlm_convergence}
\centering
\small
\setlength{\tabcolsep}{4pt}
\begin{tabular}{@{}llccccc@{}}
\toprule
\textbf{Encoding} & \textbf{Vocab} & \textbf{Ep.\,6} & \textbf{Ep.\,12} & \textbf{Ep.\,18} & \textbf{Ep.\,24} & \textbf{Ep.\,30} \\
\midrule
\beat{} (Abs, Sep)  & 186    & 14.98 & 1.70  & 1.46 & 1.40 & 1.39 \\
\beat{} (Rel, Sep)  & 185    & 8.77  & 1.47  & 1.33 & 1.30 & 1.29 \\
\beat{} (Rel, Bun)  & 7{,}145 & 60.08 & 2.47  & 1.77 & 1.66 & 1.63 \\
\beat{} (Abs, Bun)  & 7{,}145 & 86.35 & 20.81 & 2.29 & 1.88 & 1.83 \\
\midrule
REMI                & 284    & 16.49 & 2.70  & 1.51 & 1.40 & 1.39 \\
Structured          & 317    & 3.61  & 2.21  & 1.74 & 1.60 & 1.57 \\
CPWord$^\ddagger$   & 306$^*$ & 544   & 509   & 504  & 500  & 501  \\
MIDI-Like$^\dagger$ & 403    & 20.54 & 11.43 & 9.06 & 8.43 & 8.18 \\
\bottomrule
\multicolumn{7}{@{}p{\columnwidth}@{}}{\footnotesize $^*$CPWord uses 5 compound sub-vocabs ($6{\times}38{\times}156{\times}37{\times}69$); PPL reflects joint prediction}\\
\multicolumn{7}{@{}p{\columnwidth}@{}}{\footnotesize space and is not comparable to flat-vocabulary encodings.}\\
\multicolumn{7}{@{}p{\columnwidth}@{}}{\footnotesize $^\dagger$BERT MLM converged to substantially higher PPL (8.18 vs.\ ${\sim}$1.5).}\\
\multicolumn{7}{@{}p{\columnwidth}@{}}{\footnotesize $^\ddagger$CPWord PPL reflects compound vocabulary (see text).}
\end{tabular}
\end{table}

All four \beat{} schemes converge to PPL\,=\,1.29--1.83, with relative-separated (Scheme~B) achieving the lowest perplexity (1.29). Bundled schemes (C/D) start from much higher initial PPL due to their $7{,}145$-token vocabulary but undergo a sharp phase transition around epoch 12--18, ultimately converging within $0.2$--$0.5$ of separated schemes. Among alternative encodings, REMI and Structured converge normally (PPL\,=\,1.39, 1.57), while MIDI-Like plateaus at 8.18---$5$--$6\times$ higher---reflecting its flat event-stream structure where interleaved NoteOn/NoteOff tokens across all 128 pitches create long-range dependencies that MLM cannot capture. CPWord's apparent PPL of ${\sim}$500 reflects its compound vocabulary structure (joint prediction over 5 factored sub-vocabularies) rather than a failure to learn; its downstream EditF1 confirms meaningful learning.

\section{Full Per-Scheme Results}\label{app:full_results}

Tables~\ref{tab:full_beat} and~\ref{tab:full_fmd} present the complete per-scheme breakdown for all method--task combinations (perturbed-only, $n=200$).
The main text (Tables~3 and~4) reports only the best-scheme result per method; these appendix tables reveal the full encoding sensitivity.
IterEdit was evaluated across all four schemes.
SeqTag cannot perform segment completion (no generation head).

\begin{table*}[t]
\caption{Beat exact match across all encoding schemes, methods, and tasks ($n=200$). Best scheme per method--task cell in \textbf{bold}; ``---'' = not evaluated or not applicable.}
\label{tab:full_beat}
\centering
\begin{tabular}{@{}l cccc cccc cccc@{}}
\toprule
& \multicolumn{4}{c}{\textbf{Error Correction}} & \multicolumn{4}{c}{\textbf{Accomp.\ Editing}} & \multicolumn{4}{c}{\textbf{Segment Completion}} \\
\cmidrule(lr){2-5}\cmidrule(lr){6-9}\cmidrule(lr){10-13}
\textbf{Method} & A & B & C & D & A & B & C & D & A & B & C & D \\
\midrule
No-Edit      & .030 & .030 & .030 & .030 & .037 & .037 & .037 & .037 & .000 & .000 & .000 & .000 \\
Copy-Ctx & .123 & .129 & .097 & .097 & .127 & .127 & .094 & .094 & .116 & .067 & .002 & .002 \\
BERT-CMLM    & .240 & .222 & .327 & .388 & .134 & .146 & .186 & .188 & .111 & .050 & .000 & .000 \\
\midrule
SeqTag  & .700 & \textbf{.726} & .702 & .700 & .382 & .394 & \textbf{.408} & .399 & --- & --- & --- & --- \\
IterEdit & \textbf{.719} & .070 & .412 & .569 & .458 & .069 & .352 & \textbf{.480} & \textbf{.120} & .060 & .011 & .005 \\
TagFill   & \textbf{.368} & .061 & .335 & .329 & .321 & .089 & .301 & \textbf{.350} & \textbf{.436} & .264 & .298 & .326 \\
\midrule
Diffusion    & .093 & .048 & .340 & \textbf{.390} & .076 & .075 & \textbf{.200} & .198 & .053 & .050 & .080 & .080 \\
AR-Prompt & .091 & .069 & .091 & .093 & \textbf{.095} & .087 & .089 & .084 & \textbf{.152} & .105 & .044 & .043 \\
AR-Detect & \textbf{.394} & .194 & .306 & .264 & \textbf{.278} & .172 & .239 & .212 & --- & --- & --- & --- \\
AR-Select.$^\dagger$ & \textbf{.477} & .216 & .216 & .171 & \textbf{.499} & .262 & .236 & .203 & \textbf{.397} & .068 & .240 & .146 \\
Anticipatory$^\ddagger$ & --- & --- & --- & --- & --- & --- & --- & --- & .069 & --- & --- & --- \\
\bottomrule
\multicolumn{13}{@{}l@{}}{\footnotesize $^\ddagger$Anticipatory operates on raw MIDI; Scheme~A results shown for reference (encoding-independent).}
\end{tabular}
\end{table*}

\begin{table*}[t]
\caption{Fr\'echet Music Distance (FMD, $\downarrow$) across all encoding schemes, methods, and tasks ($n=200$). Best scheme per method--task cell in \textbf{bold}.}
\label{tab:full_fmd}
\centering
\begin{tabular}{@{}l cccc cccc cccc@{}}
\toprule
& \multicolumn{4}{c}{\textbf{Error Correction}} & \multicolumn{4}{c}{\textbf{Accomp.\ Editing}} & \multicolumn{4}{c}{\textbf{Segment Completion}} \\
\cmidrule(lr){2-5}\cmidrule(lr){6-9}\cmidrule(lr){10-13}
\textbf{Method} & A & B & C & D & A & B & C & D & A & B & C & D \\
\midrule
No-Edit      & 2.89 & 5.33 & 6.79 & \textbf{2.69} & 2.83 & 3.74 & 4.50 & \textbf{2.48} & 3.05 & \textbf{1.04} & 1.76 & 1.95 \\
Copy-Ctx & 4.14 & 7.74 & 9.28 & \textbf{3.90} & 3.80 & 5.48 & 6.97 & \textbf{3.62} & \textbf{2.62} & 3.80 & 3.13 & 2.81 \\
BERT-CMLM    & \textbf{0.88} & 2.05 & 2.86 & 1.01 & 1.22 & 1.50 & 1.32 & \textbf{0.97} & \textbf{2.22} & 3.88 & 2.53 & 2.35 \\
\midrule
SeqTag  & \textbf{0.08} & 0.15 & 0.32 & 0.09 & 1.86 & 1.87 & 2.56 & \textbf{1.58} & --- & --- & --- & --- \\
IterEdit & \textbf{1.03} & 8.20 & 4.09 & 1.40 & 1.52 & 3.52 & 2.00 & \textbf{1.19} & \textbf{0.34} & 1.04 & 2.49 & 2.39 \\
TagFill   & 2.03 & 3.81 & 4.02 & \textbf{1.78} & 0.99 & 2.53 & \textbf{0.83} & 1.24 & \textbf{0.27} & 0.52 & 0.42 & 0.34 \\
\midrule
Diffusion    & 14.1 & 17.4 & 2.03 & \textbf{0.87} & 18.20 & 22.08 & 2.75 & \textbf{1.72} & \textbf{14.9} & 19.6 & 52.7 & 37.5 \\
AR-Prompt & \textbf{5.56} & 11.8 & 14.2 & 11.2 & \textbf{3.89} & 6.42 & 7.14 & 6.90 & \textbf{1.54} & 7.17 & 6.65 & 8.23 \\
AR-Detect & \textbf{1.24} & 3.97 & 3.34 & 1.62 & 1.73 & 3.13 & 2.33 & \textbf{1.56} & --- & --- & --- & --- \\
AR-Select.$^\dagger$ & \textbf{1.17} & 3.79 & 4.93 & 2.36 & \textbf{0.11} & 1.93 & 1.79 & 1.18 & \textbf{0.39} & 1.20 & 0.70 & 1.03 \\
Anticipatory$^\ddagger$ & --- & --- & --- & --- & --- & --- & --- & --- & 5.69 & --- & --- & --- \\
\bottomrule
\end{tabular}
\end{table*}

Several patterns emerge.
\emph{Encoding sensitivity}: Scheme~B remains the weakest for TagFill and IterEdit, but SeqTag now achieves its highest correction score on Scheme~B (0.726), surpassing all other schemes. TagFill and IterEdit still exhibit sharp B-column drops on correction (0.061, 0.070) and editing (0.089, 0.069), suggesting that multi-pass and generative methods are more sensitive to relative-separated encoding than the single-pass classification approach of SeqTag.
The optimal encoding is method-dependent: SeqTag favors Scheme~B for correction (0.726) and Scheme~C for editing (0.408), TagFill favors Scheme~A for correction (.368) and completion (.436) but Scheme~D for editing (.350), and IterEdit is task-dependent (A for correction, D for editing).
\emph{IterEdit encoding sensitivity}: IterEdit's correction on Scheme~A (.719) far exceeds Scheme~D (.569) and~C (.412), a 0.307 gap---the largest encoding effect observed---yet for editing the pattern reverses (D: .480 $>$ A: .458).

\section{Ablation Studies}\label{app:ablation}

\subsection{IterEdit Architecture and Training Ablation}

Table~\ref{tab:levt_variants} compares IterEdit training configurations and architectural ablations on editing and correction (perturbed-only, $n=200$).
The ``full'' variants use the multi-level perturbation strategy (Section~3.3 of the main paper); ``vanilla'' uses the original contiguous-masking configuration; ``Enc-Dec'' replaces the encoder-only architecture with an encoder-decoder cross-attention design.

\begin{table}[ht]
\caption{IterEdit variant comparison (beat exact match / FMD, $n=200$). Values may differ slightly from the full per-scheme results due to different perturbation configurations used in ablation runs.}
\label{tab:levt_variants}
\centering
\begin{tabular}{@{}llcc@{}}
\toprule
\textbf{Variant} & \textbf{Sch.} & \textbf{Editing} & \textbf{Correction} \\
\midrule
Full (multi-level perturb.)   & D & \textbf{.480}\,/\,\textbf{1.19} & .568\,/\,1.40 \\
Full (multi-level perturb.)   & A & .448\,/\,2.52 & \textbf{.719}\,/\,\textbf{1.03} \\
Full + track-aware bias       & A & .428\,/\,2.54 & .666\,/\,1.10 \\
Full (multi-level perturb.)   & C & .341\,/\,2.61 & .412\,/\,4.09 \\
\midrule
vanilla (thr.\,=\,0.3)      & D & .114\,/\,2.71 & --- \\
vanilla (edit mode)          & D & --- & .129\,/\,2.27 \\
Enc-Dec (cross-attn)        & D & .051\,/\,18.5 & .034\,/\,24.3 \\
\bottomrule
\end{tabular}
\end{table}

Three ablation findings are notable.
First, adding track-aware attention bias \emph{degrades} both editing ($0.448 \to 0.428$, $-4.5\%$) and correction ($0.719 \to 0.666$, $-7.4\%$), confirming that shared attention already captures inter-track dependencies without explicit structural bias.
Second, the $4\times$ gap between full and vanilla configurations on editing beat (.480 vs.\ .114) demonstrates that multi-level perturbation training is essential for IterEdit to generalize beyond contiguous masking to fine-grained editing.
Third, replacing the encoder-only architecture with an encoder-decoder design (Enc-Dec, 66.5M vs.\ 33.9M parameters) \emph{collapses} performance (editing: .051, correction: .034) with catastrophic FMD ($>$18), confirming that the encoder-only inductive bias---where edited and context tokens share attention---is critical for the editing paradigm.

\subsection{Pre-training Ablation}\label{app:pretrain_ablation}

Table~\ref{tab:pretrain_ablation} isolates the contribution of BERT pre-training by comparing three SeqTag configurations on Scheme~D ($n=200$, perturbed-only): the full pipeline (BERT pre-trained + fine-tuned), S1-only (BERT pre-trained, no task-specific S3 fine-tuning), and Scratch (randomly initialized, no pre-training).

\begin{table}[ht]
\caption{SeqTag pre-training ablation (Scheme~D, $n=200$, perturbed-only). \textbf{Bold}: best per task.}
\label{tab:pretrain_ablation}
\centering
\begin{tabular}{@{}lcccc@{}}
\toprule
\textbf{Variant} & \textbf{Task} & \textbf{beat}$\uparrow$ & \textbf{nf1}$\uparrow$ & \textbf{FMD}$\downarrow$ \\
\midrule
Full (BERT + fine-tune) & Corr. & .700 & .852 & 0.09 \\
S1-only (BERT, no S3)   & Corr. & \textbf{.719} & \textbf{.861} & \textbf{0.07} \\
Scratch (random init)    & Corr. & .195 & .660 & 1.45 \\
\midrule
Full (BERT + fine-tune) & Edit. & .390 & .537 & 2.34 \\
S1-only (BERT, no S3)   & Edit. & \textbf{.401} & \textbf{.540} & \textbf{2.27} \\
Scratch (random init)    & Edit. & .157 & .451 & --- \\
\bottomrule
\end{tabular}
\end{table}

Two findings emerge. First, removing BERT pre-training causes catastrophic degradation: Scratch achieves only 0.195 correction beat ($-72\%$ vs.\ Full) and 0.157 editing beat ($-60\%$), confirming that domain-specific pre-training---not the SeqTag architecture itself---is the primary driver of performance. Second, the S1-only variant slightly \emph{exceeds} the full pipeline (correction: 0.719 vs.\ 0.700; editing: 0.401 vs.\ 0.390), suggesting that task-specific fine-tuning provides marginal benefit when the pre-trained representations are already well-suited to the editing task. Combined with the IterEdit encoder-decoder collapse above, these results identify two critical design choices: (1)~domain-specific BERT pre-training, and (2)~encoder-only architecture with shared attention over edited and context tokens.

\subsection{Per-Error-Type Analysis}

\begin{figure}[ht]
  \centering
  \includegraphics[width=\columnwidth]{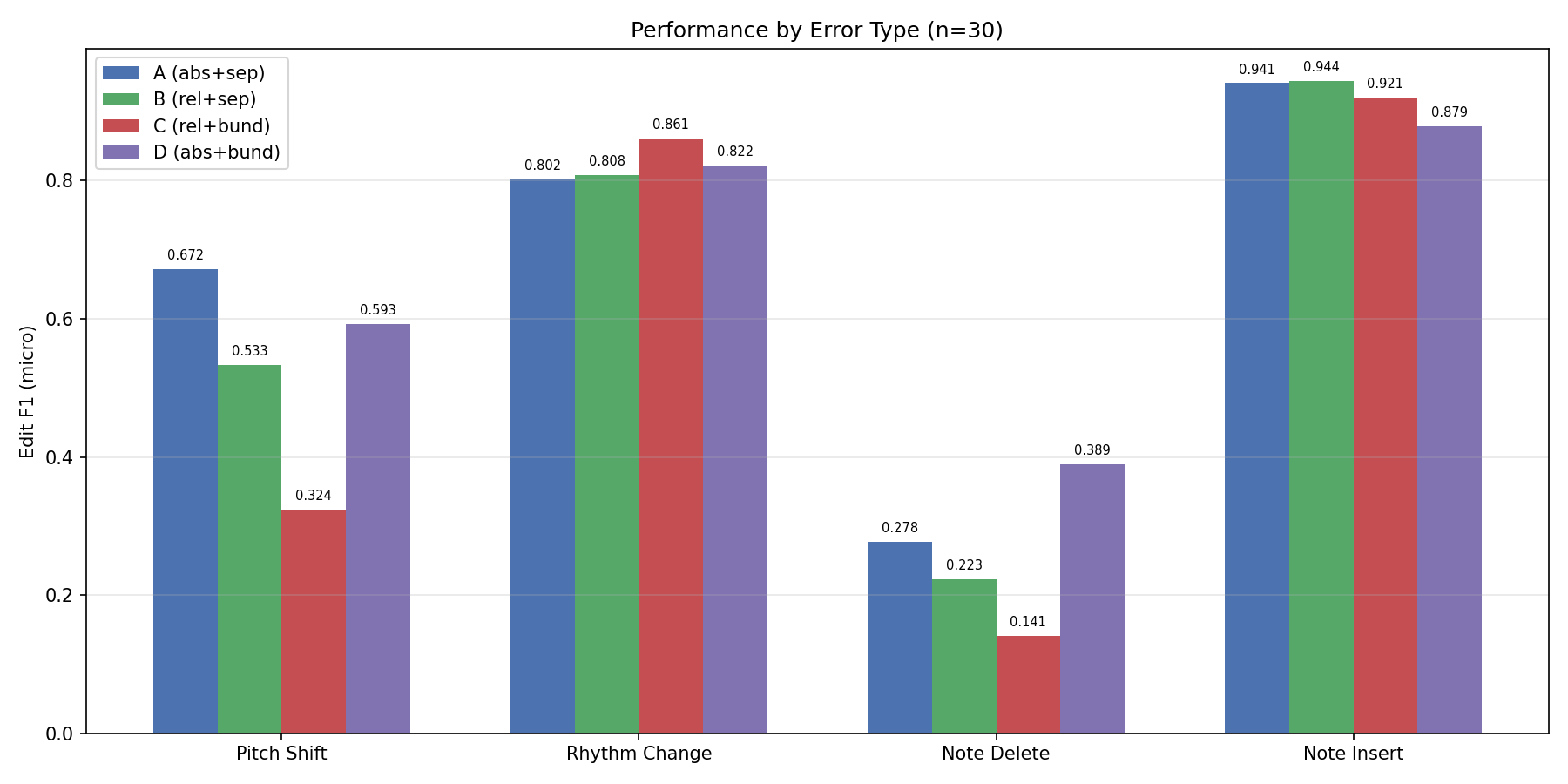}
  \caption{SeqTag correction Edit F1 by error type across encoding schemes ($n=30$ per type).}
  \label{fig:error_type}
\end{figure}

Figure~\ref{fig:error_type} breaks down SeqTag correction performance by perturbation type across encoding schemes. Note insertion is easiest (Edit F1 $>$ 0.87 for all schemes), as inserted notes create distinctive token patterns that the tagger readily detects. Rhythm changes are also well-handled (0.80--0.86). Pitch shifts are harder (0.32--0.67), especially for bundled Scheme~C where a single-token pitch change requires selecting from the full 7K vocabulary. Note deletion is the most challenging (0.14--0.39), as the tagger must identify missing content from context alone.

\subsection{Sequence Length Effect}

\begin{figure}[ht]
  \centering
  \includegraphics[width=\columnwidth]{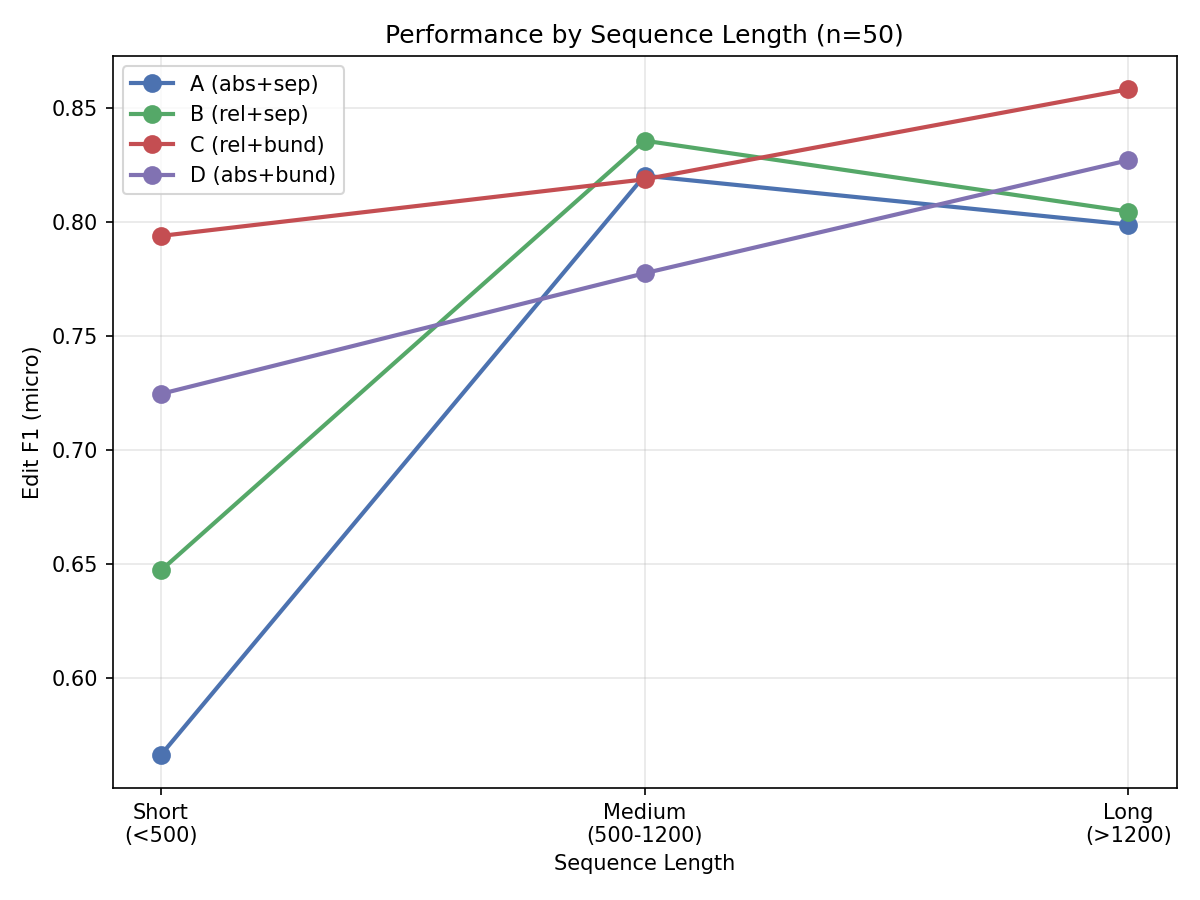}
  \caption{SeqTag correction Edit F1 by sequence length across encoding schemes ($n=50$ per bin).}
  \label{fig:length}
\end{figure}

Figure~\ref{fig:length} shows how sequence length affects correction performance. Short sequences ($<$500 tokens) are harder for separated encodings (A, B), likely due to limited context. Performance stabilizes for medium and long sequences across all schemes, with bundled Scheme~C maintaining the highest Edit F1 throughout.

\section{Statistical Significance Tests}\label{app:significance}

Pairwise significance tests use paired bootstrap resampling ($B = 10{,}000$) and Wilcoxon signed-rank tests on beat exact match ($n = 200$, perturbed-only). Effect sizes are reported as Cohen's $d$ (S: small, M: medium, L: large).

\subsection{Pairwise Method Comparisons}

\begin{table*}[t]
\caption{Key pairwise comparisons (beat exact match, 10{,}000 bootstrap resamples). Significance at $\alpha = 0.05$.}
\label{tab:pairwise_sig}
\centering
\begin{tabular}{@{}llccccc@{}}
\toprule
\textbf{Comparison (A vs.\ B)} & \textbf{Task} & \textbf{Sch.} & \textbf{Mean A} & \textbf{Mean B} & $\boldsymbol{\Delta}$ & $\boldsymbol{p}$ \\
\midrule
IterEdit\textsuperscript{A} vs.\ SeqTag\textsuperscript{B} & Corr. & --- & .719 & .726 & $-.007$ & .489 \\
IterEdit vs.\ SeqTag & Edit. & A & .458 & .382 & +.076 & $<$.001 \\
IterEdit vs.\ SeqTag & Edit. & D & .480 & .399 & +.081 & $<$.001 \\
SeqTag vs.\ TagFill & Corr. & A & .700 & .368 & +.332 & $<$.001 \\
SeqTag vs.\ TagFill & Corr. & D & .700 & .329 & +.371 & $<$.001 \\
IterEdit vs.\ TagFill & Edit. & A & .458 & .321 & +.137 & $<$.001 \\
IterEdit vs.\ TagFill & Edit. & D & .480 & .350 & +.130 & $<$.001 \\
TagFill vs.\ No-Edit & Compl. & A & .436 & .000 & +.436 & $<$.001 \\
TagFill vs.\ No-Edit & Compl. & D & .326 & .000 & +.326 & $<$.001 \\
\bottomrule
\end{tabular}
\end{table*}

The key finding is that IterEdit and SeqTag show no statistically significant difference on correction at their respective best schemes ($p = 0.489$), while IterEdit is \emph{significantly} better on editing in Schemes~A and~D ($p < 0.001$). All other pairwise comparisons in Table~\ref{tab:pairwise_sig} are also highly significant ($p < 0.001$).

\subsection{Encoding $\times$ Method Interaction (ANOVA)}

\begin{table}[htbp]
\caption{Two-way ANOVA: encoding $\times$ method interaction on token-space beat exact match ($n=50$ per cell). Correction includes SeqTag, IterEdit, and TagFill; editing additionally includes No-Edit. All interactions are significant at $\alpha = 0.05$.}
\label{tab:anova}
\centering
\small
\begin{tabular}{@{}lcccc@{}}
\toprule
\textbf{Analysis} & \textbf{Factor} & $\boldsymbol{F}$ & $\boldsymbol{p}$ & $\boldsymbol{\eta^2}$ \\
\midrule
\multirow{3}{*}{\shortstack[l]{Corr.: 3 methods\\$\times$ 4 enc (A,B,C,D)}}
  & Method   & 271.6 & $<$0.001 & 34.6\% \\
  & Encoding & 80.4  & $<$0.001 & 15.4\% \\
  & Interaction & 32.7 & $<$0.001 & 12.5\% \\
\midrule
\multirow{3}{*}{\shortstack[l]{Edit.: 4 methods\\$\times$ 4 enc (A,B,C,D)}}
  & Method   & 68.8  & $<$0.001 & 18.7\% \\
  & Encoding & 18.4  & $<$0.001 & 5.0\% \\
  & Interaction & 6.3 & $<$0.001 & 5.1\% \\
\bottomrule
\end{tabular}
\end{table}

All encoding$\times$method interactions are highly significant ($p < 0.001$), formally confirming that method rankings reverse across encoding schemes. The interaction effect is largest for correction ($\eta^2 = 12.5\%$), driven primarily by TagFill and IterEdit's sharp drops on Scheme~B (0.061, 0.070) while SeqTag achieves its highest score on B (0.726). For editing, the interaction ($\eta^2 = 5.1\%$) reflects TagFill's Scheme~B collapse (0.084) versus its competitive performance on other schemes. Encoding explains substantially more variance in correction (15.4\%) than in editing (5.0\%), consistent with the main text finding that encoding is a stronger design lever for precision-demanding tasks.

\section{Subjective Evaluation Details}\label{app:subjective}

We recruited 33 evaluators from university students (34\% no music training, 42\% hobby-level, 18\% formally trained 3+ years, 6\% music majors; 55\% play an instrument; 52\% listen daily). Systems were presented in randomized blind order via a web interface with piano roll visualizations and audio playback. Each evaluator provided 228 ratings across four experiments on three dimensions: music quality (MQ), naturalness (NA), and edit accuracy (EA), using a 1--5 MOS scale. Tables~\ref{tab:sub_enc}--\ref{tab:sub_edit} report per-dimension scores.

\textbf{Reliability and significance.} Inter-rater agreement, computed as average-measures ICC(2,k), ranges from 0.75 to 0.95 across the four experiments (mean 0.87), i.e.\ good to excellent. For each experiment and dimension, a Friedman omnibus test rejects the null hypothesis of equal system ratings (all $p \leq .001$). Holm-corrected pairwise Wilcoxon signed-rank tests confirm the main comparisons: \beat{}-C outperforms all alternative encodings ($p < .001$); on error correction, SeqTag and IterEdit outperform all generative baselines ($p < .05$); and TagFill leads on segment completion ($p < .01$).

\begin{table}[htbp]
\caption{Exp.~1: Error correction --- encoding comparison.}
\label{tab:sub_enc}
\centering
\begin{tabular}{@{}lcccc@{}}
\toprule
\textbf{System} & \textbf{MQ} & \textbf{NA} & \textbf{EA} & \textbf{Avg} \\
\midrule
\beat{}-C + SeqTag & \textbf{4.19} & \textbf{4.23} & \textbf{4.29} & \textbf{4.24} \\
REMI + SeqTag & 2.99 & 2.64 & 2.78 & 2.80 \\
CPWord + SeqTag & 2.81 & 2.60 & 2.67 & 2.69 \\
Structured + SeqTag & 2.51 & 2.35 & 2.47 & 2.45 \\
\bottomrule
\end{tabular}
\end{table}

\begin{table}[htbp]
\caption{Exp.~2: Error correction --- method comparison.}
\label{tab:sub_corr}
\centering
\begin{tabular}{@{}lcccc@{}}
\toprule
\textbf{System} & \textbf{MQ} & \textbf{NA} & \textbf{EA} & \textbf{Avg} \\
\midrule
SeqTag & 4.18 & \textbf{4.23} & \textbf{4.23} & \textbf{4.21} \\
IterEdit & \textbf{4.14} & 4.15 & 4.09 & 4.13 \\
TagFill & 3.32 & 3.55 & 3.64 & 3.51 \\
Diffusion SDEdit & 3.40 & 3.23 & 3.11 & 3.25 \\
BERT-CMLM & 3.37 & 3.06 & 3.31 & 3.25 \\
AR-Detect & 2.95 & 2.99 & 3.19 & 3.05 \\
\bottomrule
\end{tabular}
\end{table}

\begin{table}[htbp]
\caption{Exp.~3: Segment completion.}
\label{tab:sub_compl}
\centering
\begin{tabular}{@{}lcccc@{}}
\toprule
\textbf{System} & \textbf{MQ} & \textbf{NA} & \textbf{EA} & \textbf{Avg} \\
\midrule
TagFill & \textbf{4.10} & \textbf{4.17} & \textbf{4.17} & \textbf{4.15} \\
IterEdit & 3.25 & 3.45 & 3.56 & 3.42 \\
LLaMA-Selective & 3.38 & 3.02 & 3.07 & 3.15 \\
Anticipatory & 3.05 & 2.68 & 2.86 & 2.86 \\
\bottomrule
\end{tabular}
\end{table}

\begin{table}[htbp]
\caption{Exp.~4: Accompaniment editing.}
\label{tab:sub_edit}
\centering
\begin{tabular}{@{}lcccc@{}}
\toprule
\textbf{System} & \textbf{MQ} & \textbf{NA} & \textbf{EA} & \textbf{Avg} \\
\midrule
IterEdit & \textbf{3.96} & \textbf{4.14} & \textbf{4.12} & \textbf{4.07} \\
SeqTag & 3.79 & 3.92 & 4.03 & 3.91 \\
TagFill & 3.80 & 3.89 & 3.80 & 3.83 \\
LLaMA-Selective & 3.45 & 3.08 & 3.09 & 3.21 \\
BERT-CMLM & 3.23 & 3.08 & 3.12 & 3.14 \\
Diffusion SDEdit & 3.08 & 2.84 & 2.90 & 2.94 \\
\bottomrule
\end{tabular}
\end{table}

\section{Efficiency Analysis}\label{app:efficiency}

Table~\ref{tab:latency_per_task} reports per-task latency for all methods. Single-pass methods (CMLM, SeqTag, TagFill) are stable across tasks (37--65\,ms). IterEdit varies substantially: completion requires few iterations (82\,ms) while editing requires many (655\,ms). AR latency depends on the generation strategy and output length.

\begin{table}[ht]
\caption{Per-task inference latency (ms) and parameters.}
\label{tab:latency_per_task}
\centering
\begin{tabular}{@{}lrrrrr@{}}
\toprule
\textbf{Method} & \textbf{Corr.} & \textbf{Edit.} & \textbf{Compl.} & \textbf{Params (M)} \\
\midrule
BERT-CMLM      & 40   & 42    & 37    & 30.2 \\
SeqTag          & 65   & 57    & ---   & 37.2 \\
TagFill         & 60   & 60    & 60    & 63.8 \\
IterEdit        & 273  & 655   & 82    & 33.9 \\
Diffusion r03   & 445  & 484   & 434   & 34.4 \\
Anticipatory    & ---  & ---   & 409   & 128.1 \\
AR-Selective    & 9{,}357  & ---   & ---   & 162.0 \\
AR-Detect       & 9{,}677  & 9{,}458 & ---   & 199.2 \\
AR-Teacher      & 12{,}634 & ---   & ---   & 162.0 \\
AR-Prompt       & 22{,}934 & ---   & 23{,}654 & 162.0 \\
\bottomrule
\end{tabular}
\end{table}

\textbf{Architecture-agnostic FLOPs.} Wall-clock latency is affected by implementation and model size. For an implementation-independent view, Table~\ref{tab:flops} reports analytically estimated single-sample inference FLOPs (standard transformer accounting, equal sequence length 2{,}048). Editing methods complete inference in 1--3 forward passes (172--517 GFLOPs). Notably, AR totals only 789 GFLOPs thanks to KV caching, yet is two orders of magnitude slower in wall-clock time because its 2{,}048 decoding steps must execute strictly sequentially; D3PM requires 30 denoising passes (5{,}604 GFLOPs). The efficiency advantage of editing is thus paradigmatic---few parallel passes---rather than an artifact of implementation-level optimization.

\begin{table}[ht]
\caption{Estimated single-sample inference FLOPs at equal sequence length 2{,}048. Forward passes are the dominant factor.}
\label{tab:flops}
\centering
\begin{tabular}{@{}lrr@{}}
\toprule
\textbf{Method} & \textbf{Forward passes} & \textbf{GFLOPs} \\
\midrule
SeqTag           & 1                    & 172 \\
BERT-CMLM        & 1                    & 172 \\
TagFill          & 3 (1 tag + 2 fill)   & 517 \\
IterEdit         & ${\sim}$3            & 517 \\
Diffusion (D3PM) & 30                   & 5{,}604 \\
AR (LLaMA)       & 2{,}048 (sequential) & 789 \\
\bottomrule
\end{tabular}
\end{table}

We further examine how encoding scheme affects inference speed within the SeqTag pipeline.

\subsection{Per-Encoding Efficiency}

Table~\ref{tab:efficiency} further examines how encoding scheme affects inference speed within the SeqTag pipeline ($n \approx 20$ samples per scheme, seed~42).

\begin{table}[ht]
\caption{Inference efficiency by encoding scheme (SeqTag pipeline).}
\label{tab:efficiency}
\centering
\begin{tabular}{@{}lcccc@{}}
\toprule
\textbf{Metric} & \textbf{A} & \textbf{B} & \textbf{C} & \textbf{D} \\
\midrule
Time / sample (ms)  & 51.4  & 54.9  & 38.3  & \textbf{37.6} \\
Samples / sec       & 19.5  & 18.2  & 26.1  & \textbf{26.6} \\
Mean iterations      & 2.12  & 2.11  & \textbf{2.00}  & \textbf{2.00} \\
p95 latency (ms)    & 83.5  & 112.3 & 72.3  & \textbf{71.8} \\
\midrule
Model memory (MB)    & \textbf{101}   & \textbf{101}   & 142   & 142 \\
Peak memory (MB)     & \textbf{203}   & 211   & 380   & 380 \\
\bottomrule
\end{tabular}
\end{table}

Bundled encodings (C, D) are 35--40\% faster than separated encodings (A, B) because bundled beats produce shorter token sequences, reducing both self-attention cost and iteration count (converging in exactly 2 iterations vs.\ a mean of 2.1 for separated).
The p95 tail latency of Scheme~B (112\,ms) is notably higher than other schemes, consistent with its longer average token length (1\,445 vs.\ 1\,096--1\,382 tokens).
Memory usage is higher for bundled schemes ($\sim$380\,MB peak vs.\ $\sim$200\,MB) due to the larger vocabulary, but all configurations remain well within single-GPU capacity.
Combined with the cross-method results above, the overall picture is clear: edit-based methods converge in 2--3 constant-time iterations regardless of sequence length, with bundled encoding providing an additional speed advantage at the cost of higher memory.

\section{Multi-Track Generalization Details}\label{app:multi_track}

The multi-track dataset comprises 108,055 pieces from the Lakh MIDI Dataset, covering 105 General MIDI programs (Piano 19.0\%, Drums 13.3\%, Electric Bass 4.5\%, Strings 4.4\%, Guitar 11.1\% combined). Pieces contain 2--10 tracks (mean 6.2, median 6) and 17--243 measures (mean 86.6). Tracks are merged into two voices and encoded with Scheme~C ($|V|{=}7{,}145$), with 64.6\% of sequences exceeding the 2,048-token limit and requiring truncation at bar boundaries.

Table~\ref{tab:multi_track_full} provides per-task breakdowns. SeqTag on correction achieves 0.258 beat exact match with notably high nf1 (0.797) and low MPE (3.57), suggesting that multi-track accompaniment has more predictable pitch patterns than piano. TagFill on completion reaches 0.271 beat exact match with FMD of 0.49---the lowest among all task--setting combinations---indicating high distributional quality. IterEdit on editing shows a beat drop (0.269$\to$0.196) with comparable nf1 (0.435$\to$0.534).

\begin{table}[ht]
\caption{Multi-track generalization: full metrics per task (Scheme~C, perturbed beats only). Piano values under the same Scheme~C for reference.}
\label{tab:multi_track_full}
\centering
\begin{tabular}{@{}llccc@{}}
\toprule
\textbf{Task} & \textbf{Setting} & \textbf{beat}$\uparrow$ & \textbf{nf1}$\uparrow$ & \textbf{cf1}$\uparrow$ \\
\midrule
\multirow{4}{*}{Corr.}
  & Piano No-Edit & .035 & .435 & .452 \\
  & Piano SeqTag  & .400 & .531 & .546 \\
  & Multi No-Edit & .000 & .803 & .798 \\
  & Multi SeqTag  & .258 & .797 & .804 \\
\midrule
\multirow{4}{*}{Compl.}
  & Piano No-Edit & .000 & .000 & .000 \\
  & Piano TagFill & .298 & .442 & .507 \\
  & Multi No-Edit & .028 & .000 & .000 \\
  & Multi TagFill & .271 & .494 & .628 \\
\midrule
\multirow{4}{*}{Edit.}
  & Piano No-Edit & .035 & .435 & .452 \\
  & Piano IterEdit & .269 & .435 & .448 \\
  & Multi No-Edit & .003 & .539 & .565 \\
  & Multi IterEdit & .196 & .534 & .568 \\
\bottomrule
\end{tabular}
\end{table}

A consistent pattern across all three tasks: beat exact match is lower on multi-track data (reflecting greater diversity), while note-level metrics (nf1, chroma\_f1) are comparable or better (reflecting pitch redundancy in ensemble arrangements). The framework transfers without architectural modification, confirming that the \beat{} encoding's structural properties generalize beyond the piano domain.

\end{document}